\documentclass[apj]{emulateapj}
\usepackage{epsf}
\usepackage{graphicx,color}

\newcommand{\lsim}{\lower0.6ex\vbox{\hbox{$ \buildrel{\textstyle <}\over{\sim}\ $}}}
\newcommand{\gsim}{\lower0.6ex\vbox{\hbox{$ \buildrel{\textstyle >}\over{\sim}\ $}}}

\newcommand{\pink}{\color{\pink}}

\newcommand{\hkpc}{h^{-1}\mathrm{kpc}}

\newcommand{\msun}{\mathrm{M}_{\odot}}

\newcommand{\hMpc}{h^{-1}\ \mathrm{Mpc}}

\newcommand{\kms}{{\,{\rm km\ s}^{-1}}}

\newcommand{\kpc}{{\,{\rm kpc}}}
\newcommand{\Mpc}{{\,{\rm Mpc}}}

\newcommand{\lcdm}{\Lambda\mbox{CDM}}

\newcommand{\lw}{\ell}
\newcommand{\lwmin}{\lw_{min}}

\newcommand{\avg}[1]{\langle #1 \rangle}

\newcommand{\lr}{\lambda_{\mathrm{R}}}
\newcommand{\lt}{\lambda_{\mathrm{T}}}

\newcommand{\bm}[1]{\mathbf{#1}}

\newcommand{\na}{n_{\rm arcs}}
\newcommand{\bp}{\bm{p}}
\newcommand{\ns}{n_s}

\begin{document}

\submitted{The Astrophysical Journal, submitted}
\vspace{1mm}
\slugcomment{{\em The Astrophysical Journal, submitted}}

\shortauthors{Rozo et al.}
\shorttitle{Impact of Baryonic Cooling on Giant Arcs}

\title{The Impact of Baryonic Cooling on Giant Arc Abundances}

\author{
Eduardo Rozo\altaffilmark{1,2,3},
Daisuke Nagai\altaffilmark{4},	
Charles Keeton\altaffilmark{5},
Andrey Kravtsov\altaffilmark{2,6,7}
}


\begin{abstract}

  Using ray tracing for simple analytic profiles, we demonstrate that
  the lensing cross section for producing giant arcs has distinct
  contributions due to arcs formed through image distortion only, and
  arcs form from the merging of two or three images.  We investigate
  the dependence of each of these contributions on halo ellipticity
  and on the slope of the density profile, and demonstrate that at
  fixed Einstein radius, the lensing cross section increases as the
  halo profile becomes steeper.  We then compare simulations with and
  without baryonic cooling of the same cluster for a sample of six
  clusters, and demonstrate that cooling can increase the overall
  abundance of giant arcs by factors of a few.  The net boost to the
  lensing probability for individual clusters is mass dependent, and
  can lower the effective low mass limit of lensing clusters.  This
  last effect can potentially increase the number of lensing clusters
  by an extra $50\%$.  While the magnitude of these effects may be
  overestimated due to the well known overcooling problem in
  simulations, it is evident that baryonic cooling has a
  non-negligible impact on the expected abundance of giant arcs, and
  hence cosmological constraints from giant arc abundances may be
  subject to large systematic errors.
\end{abstract}


\keywords{cosmology: theory -- lensing -- giant arcs}


\altaffiltext{1}{
Department of Physics, The University of Chicago, 
Chicago, IL 60637, USA
}

\altaffiltext{2}{
Kavli Institute for Cosmological Physics, 
Chicago, IL 60637, USA
}

\altaffiltext{3}{
CCAPP Postdoctoral Fellow, Department of Physics, 
The Ohio State University,
Columbus, OH 43210, USA
{\tt erozo@mps.ohio-state.edu}
}

\altaffiltext{4}{
 Theoretical Astrophysics,
California Institute of Technology,
Mail Code 130-33,
Pasadena, CA 91125 
}

\altaffiltext{5}{
Department of Physics and Astronomy,
Rutgers University,
136 Frelinghuysen Road,
Piscataway, NJ 08854, USA
}

\altaffiltext{6}{
Department of Astronomy and Astrophysics, 
The University of Chicago, 
Chicago, IL 60637, USA
}

\altaffiltext{7}{
Enrico Fermi Institute, 
The University of Chicago, 
Chicago, IL 60637, USA
}


\section{Introduction}

Soon after the discovery of giant arcs
\citep[][]{lyndspetrosian86,soucailetal87} and their subsequent
interpretation as gravitational lensing events
\citep[][]{paczynski87}, \citet[][]{grossmannarayan88} realized that
the statistics of these relatively rare events can be used as a
cosmological probe.  This idea was greatly expanded by
\citet[][]{miraldaescude93} and \citet[][]{wuhammer93} who used arc
statistics to constrain the mass distribution of clusters and the high
redshift galaxy population
\citep[][]{miraldaescude93b,bezecourtetal98}.  Attempts to derive
cosmological parameters from arc abundances soon followed, and
concluded that giant arc abundances had a relatively mild
cosmological dependence \citep[][]{wumao96}.

These early studies used simple analytical models of cluster mass
distributions, which turn out to severely under-predict lensing
probabilities relative to clusters in numerical simulations
\citep[][]{bartelmannweiss94,bartelmannetal95}.  Nevertheless, it
seemed reasonable to expect that the {\em scaling} of arc abundances
with cosmology should be broadly consistent with predictions from
analytic prescriptions, so it was surprising when the first numerical
study to investigate arc abundances in various cosmologies
\citep[][]{bartelmannetal98} found not only that there were order of
magnitude differences in the predictions from different cosmologies,
but also that the abundance of giant arcs expected in the now-standard
$\lcdm$ cosmology is an order of magnitude lower than the observed
abundance \citep[][]{lefevreetal94}.

The claimed sensitivity of giant arc abundances to cosmology and the
strong inconsistency between theory and observations found by
\citet[][]{bartelmannetal98} generated an explosion of both
theoretical and observational studies.  On the observational front,
new arc samples with more clearly defined selection functions became
available \citep[][]{luppinoetal99,zaritskygonzalez03,gladdersetal03}.
At the same time, the cosmological community mounted an extensive
theoretical effort to search for possible systematics and/or
theoretical uncertainties in order to determine whether the
disagreement was real.

The theoretical program has revealed that the problem of giant arc
statistics is extremely rich, with predictions being sensitive to
details of both background galaxies and cluster properties.
Concerning sources, many models for giant arc abundances have assumed
circular sources, but source ellipticity can significantly boost the
number of giant arcs in the sky \cite[][]{bartelmannetal95,keeton01b}.
Assumptions about the distribution of source radii appear to be
important \citep[][]{miraldaescude93b,oguri02,howhite05}.  The
redshift distribution of sources plays a significant role
\citep[][]{wambsganssbodeostriker04,dalalholderhennawi04} and in fact
contributes much of the theoretical uncertainty in modern predictions
\citep[][]{hennawietal05}.

There are also several systematics associated with the lens
population.  For instance, while many studies have assumed spherically
symmetric mass distributions
\citep[e.g.,][]{hattorietal97,hamanafutamase97,molikawaetal99,
  cooray99,cooray99b,williamsetal99}, many others have pointed out
that halo ellipticity significantly increases giant arc abundances
\citep[e.g.,][]{grossmannarayan88,bartelmannetal95,oguri02,meneghettietal03}.
Moreover, halo triaxiality can give rise to order of magnitude
variations in the lensing probability of a given cluster with viewing
angle \citep[][]{ogurietal03,dalalholderhennawi04}, and therefore must
be treated properly in arc abundance predictions.  Lensing
probabilities for individual clusters appear to be significantly
enhanced during mergers
\citep[e.g.,][]{torrietal04,meneghettietal05b,fedelietal06}, though
such episodes appear to be rare and transient enough that the overall
lens populations is not significantly biased towards a cluster merging
population \citep[][]{hennawietal05}.

One final related trend, noted in both analytical and numerical works,
is that lensing probabilities are very sensitive to the radial density
profiles of lensing clusters
\citep[e.g.,][]{miraldaescude93,hattorietal97,ogurietal01,
  meneghettietal05,hennawietal05}.  Consequently, clear predictions
for the distribution of cluster density profiles must be in place for
arc abundance predictions to become robust.

Systematics that do {\em not} seem to have significant effects on
lensing probabilities have also been identified.  For instance,
studies in which a central galaxy is ``painted'' on the center of
simulated clusters suggest that this additional mass component has
minimal impact on the expected giant arc abundance
\citep[][]{molikawahattori01,meneghettietal03b,dalalholderhennawi04}.
Likewise, the presence of substructures and/or galaxies in the cluster
also seems to have a negligible effect on lensing probabilities
\citep[][]{grossmannarayan88,floresetal00,meneghettietal00,hennawietal05}.
Line of sight projection effects \citep[][]{wambsganssetal05} also
appear to be unimportant \citep[][]{hennawietal05}.

At present, there seems to be general agreement between theory and
observations
\citep[][]{dalalholderhennawi04,hennawietal05,horeshetal05}, although
the situation is not completely clear \citep[][]{lietal05}.  The new
agreement arises from the coherent contributions of various effects,
with no single systematic explaining the order of magnitude effects
observed by \citet[][]{bartelmannetal98}.  Recent analytical
\citep[][]{kaufmannstraumann00,bartelmannetal03} and numerical
\citep[][]{meneghettietal05} studies suggest that variations in the
arc abundance between different cosmologies are factors of a few
rather than an order of magnitude, further relaxing any real tension
between theory and observations.  What seems clear at this time is
that arc abundances are closer to old cosmology than to precision
cosmology: agreement is to be understood at the factor of two level,
even if current observations did not suffer from small number
statistics.

Given the complexity of the problem, it is important to consider
whether there are any additional systematic effects that alter
theoretical predictions.  The goal of this paper is to study one such
possibility: the impact of baryonic cooling on cluster mass profiles.
Numerical simulations that include cooling have shown that once
baryons have cooled and sunk to the center of the halo to form the
cluster's central galaxy, the dark matter halo itself contracts in
response to the increased central density
\citep[][]{blumenthaletal96,kochanekwhite01,
  gnedinetal04,sellwoodmcgaugh05}.  The effects of this contraction
are significant, and include a steepening of the inner regions of the
halo profile, a boost to the mass enclosed within a fixed radius $r$,
and an increase of the degree of spherical symmetry of the mass
distribution \citep[][]{kazantzidisetal04}.  Since arc abundances are
known to be sensitive to cluster density profiles we expect that
baryonic cooling may significantly affect the predictions.  Indeed,
recent work by \citet[][]{puchweinetal05} has shown that baryonic
cooling enhances the lensing probability of very massive clusters
($M\gtrsim 10^{15}\ \msun$) by a factor of $\sim$2--4.
\citet[][]{puchweinetal05} did not consider, however, the effect of
baryonic cooling on smaller clusters, which is particularly
significant given that the median mass of the cluster lens sample
\citep[see][]{hennawietal05} is well below that considered in
\citet[][]{puchweinetal05}.

In this work, we extend the analysis of the impact of baryonic cooling
on arc statistics to clusters of considerably smaller masses.  In
particular, we demonstrate that baryonic cooling affects lensing
probabilities for low mass clusters more than for high mass clusters,
both because low mass clusters have a larger cooled mass fraction, and
because lensing in low mass clusters is more confined to the central
regions that are strongly affected by cooling.  We also demonstrate
that baryonic cooling has a strong isotropization effect for both the
lensing probability of individual cluster as a function of viewing
angle, and the spatial distribution of arcs around the cluster center.
In order to understand {\em why} baryonic cooling can have such
significant impact on the lensing cross section, we develop a
theoretical framework that allows us to interpret the lensing
probability associated with each cluster as a function of the minimum
length to width ratio of the arcs considered.  We have found that our
theoretical framework can be used to better interpret and
contextualize results from previous works, providing an intuitive base
upon which the sensitivity of arc statistics to various cluster
parameters can be deduced in a qualitative fashion.


\section{Lensing Cross Sections}
\label{sec:cs}

Lensing probabilities are usually characterized in terms of cross
sections, so for completeness we review the concept of the lensing
cross section.  Let $\na$ denote the number of giant arcs per unit
area in the sky, expressed as,
\begin{equation}
\na = \int dz_s dz_l d\bp\ \frac{d\na}{dz_s dz_l d\bp}
\end{equation}
where $d\na/dz_s dz_l d\bp$ is the number of giant arcs due to lenses
at redshift $z_l$, given sources at redshift $z_s$ with structure
parameters $\bp$ (such as radius and ellipticity).  Our immediate goal
is to compute $d\na/dz_s dz_l d\bp$.  Let $\bm{x}$ denote a position
vector on the source shell at redshift $z_s$ behind a cluster at
redshift $z_l$.  If $r(\bm{x})$ is the length to width ratio of the
image of a source at position $\bm{x}$, then the expected number of
arcs with a length to width ratio larger than $\lw$ behind this
cluster can be written as $d\na(\lw) = d\ns \sigma(\lw;\bp)$.  Here
$d\ns$ is the number density of sources at redshift $z_s$ with
parameters $d\bp$: $d\ns = (d\ns/d\bp)(dV/dz_s)dz_s d\bp$.  Also,
$\sigma(\lw;\bp)$ is the area of the source plane over which the image
of the source satisfies $r(\bm{x})>\lw$.  This area $\sigma(\lw)$ is
called the {\em lensing cross section}, and is defined as
\begin{equation}
\sigma(\lw) = \int_{r(\bm{x})>\lw} d^2\bm{x}\ w(\bm{x};\lw)
\end{equation}
where $w(\bm{x};\lw)$ is the number of images of a source placed at
point $\bf{x}$ with a length to width ratio larger than
$\lw$.\footnote{Note that with this definition the phrase ``lensing
  cross section'' always refers to a cumulative quantity.  The
  differential lensing cross section $d\sigma/d\lw$ is the
  multiplicity-weighted area over which sources generate arcs of
  length to width ratio exactly equal to $\lw$.}  If we now denote the
average lensing cross section of mass $m$ halos as
$\avg{\sigma|m;z_s,z_l,\bp}$, it follows that the number of expected
arcs is given by
\begin{eqnarray} \label{eq:arcab}
\frac{d\na}{dz_sdz_ld\bp} &=& 
	\frac{d\ns(\bp;z_s)}{d\bp}\frac{dV}{dz_s} \\
  && \times \int dm\ \frac{dn_{\rm halos}}{dm}(m,z_l)\frac{dV}{dz_l} 
			\avg{\sigma|m;z_s,z_l,\bp},
  \nonumber
\end{eqnarray}
where $dn_{\rm halos}/dm$ is the number density of halos of mass $m$
at redshift $z_l$.  We can see that the lensing signal is fully
characterized by the average lensing cross section
$\avg{\sigma|m;z_s,z_l,\bp}$ and the halo mass function $dn/dm$.

To summarize, in order to compute the expected number of giant arcs in
the universe one needs to know:
\begin{itemize}
\item The halo abundance $dn/dm$ as a function of lens redshift $z_l$.
\item The source abundance $d\ns/d\bp$ as a function of source
  redshift $z_s$ and structure parameters $\bp$.
\item The average lensing cross section for halos of mass $m$, as a
function of lens and source redshift and source structure parameters.
\end{itemize}
Given the evident complexity of the problem, studies that analyze the
dependence of lensing cross sections on various parameters are
particularly important: they help identify aspects of the theory that
need to be well characterized to obtain robust predictions for arc
abundances.  As we shall see, the lensing cross section $\sigma(\lw)$
tends to exhibit some robust qualitative features, so in this paper we
devote considerable effort to understanding these general features in
order to provide a robust framework in which we can better interpret
our numerical results.


\section{Lensing Algorithm and Selection Function}

The idea behind our code is simple enough: simulation outputs are used
to construct three dimensional pixelized density maps of the clusters,
which are then projected to create pixelized surface density maps.
Using Fast Fourier Transforms, we compute the gravitational potential,
angular deflection, and inverse magnification tensor at each grid
point in the lens plane.  Following \citet[][]{hennawietal05}, we map
each lens plane pixel back to the source plane, where we lay down a
grid and associate each source plane pixel with all lens plane pixels
contained within it.  This yields a lookup table that allows us to
quickly find which image (lens plane) pixels are lit given a set of
lit source pixels.  Circular sources are then placed along an
additional coarser grid on the source plane, referred to as the
placement grid.  For each source, we identify its images and measure
their lengths and widths.  The lensing cross section for arcs above a
given length to width ratio is then estimated from the number of
placement pixels which generate appropriate arcs, weighted by the area
of each placement grid pixel and the number of arcs produced by said
source.  All of the various parameters relevant for the algorithm
(e.g., grid scales, grid extent, number of pixels, etc.) are chosen to
ensure that the resulting cross sections are accurate to 10\% or
better.  We note that because our goal is to estimate the {\em
  relative} change in the lensing cross section of a cluster due to
baryonic cooling, we do not expect our conclusions to be very
sensitive to assumptions about source size and shape. We have
therefore opted to make simple assumptions, and leave a detailed
analysis of how source properties affect lensing cross sections to
future work.

The details of our algorithm are presented in Appendix
\ref{app:algorithm}, including various procedures to secure a $10\%$
accuracy in the cross section, and several techniques to improve the
speed of the algorithm. In this study, we have chosen to focus on
tangential arcs only.  This is not only because interpretation of the
results is simpler when neglecting radial arcs (see Appendix
\ref{app:algorithm}, and the discussion in the following sections),
but also because we expect the observational selection function for
tangential arcs to be simpler than that of radial arcs since the
former are more prominent and reside further away from a cluster's
central galaxy than their radial counterparts.


\section{Analytic Models}
\label{sec:analytic_models}

We first study the cross section function $\sigma(\lw)$ for a variety
of simple analytical models in order to develop a general
understanding of its features.  This basic framework will be used in
section \ref{sec:simulations} to analyze the cross section function of
numerical clusters.


\subsection{The Models}
\label{sec:halomodels}

We study halos with projected density profiles of the form
\begin{equation}
\kappa = \frac{b}{2}\ \frac{f(q)h(\beta)}{(q^2x^2+y^2)^{\beta/2}}
\label{eq:profile}
\end{equation}
where $q$ is the projected axis ratio, and $b$ corresponds to the
Einstein radius of the profile when $q=1$.  The function $f(q)$ is
defined by demanding that the mass contained within a circle of radius
$b$ be independent of the axis ratio $q$.  The parameter $\beta$ is
the logarithmic slope of the profile, and $h(\beta)$ is defined by
setting $h(1)=1$ and demanding that the mass contained within a circle
of radius $b$ be independent of $\beta$.  The special case $\beta=1$
corresponds to the Singular Isothermal Ellipsoid (SIE), and the
spherically symmetric SIE case is called the Singular Isothermal
Sphere (SIS).  For an SIS, the relation between Einstein radius and
velocity dispersion is
\begin{equation}
b=4\pi\left(\frac{\sigma_v}{c}\right)^2 \frac{D_{ls}}{D_{os}}
\end{equation}
where $D_{ls}$ and $D_{os}$ are the angular diameter distances
between the lens and the source and between the observer and the
source, respectively.  

In what follows, we consider two sets of models.  For the first set,
we fix $\beta=1$ (SIE) and then consider Einstein radii corresponding
to velocity dispersions $\sigma_v/(10^3 \kms) \in
\{0.6,0.7,0.8,0.9,1.0\}$.  We also take $q \in
\{1.0,0.9,0.8,0.7,0.6,0.5,0.4\}$ for a total of 35 models.  For the
second set, we fix the Einstein radius to that of an SIS with velocity
dispersion $\sigma_v=10^3 \kms$, and then consider density slopes
$\beta \in \{0.4,0.6,0.8,1.0,1.2,1.4\}$.  We use the same set of axis
ratios, for a total of 35 additional models.  In all cases we assume
typical lens and source redshifts of $z_l=0.3$ and $z_s=1.5$.


\subsection{$\lw$ Maps and Cross-Sections}

Since the lensing cross section $\sigma(\lw)$ is simply the area of
the source plane in which source produces arcs with length to widths
larger than $\lw$, we begin our analysis by looking at $\lw$ maps for
our analytic SIE models.

\begin{figure}[t]
\epsscale{1.2}
\plotone{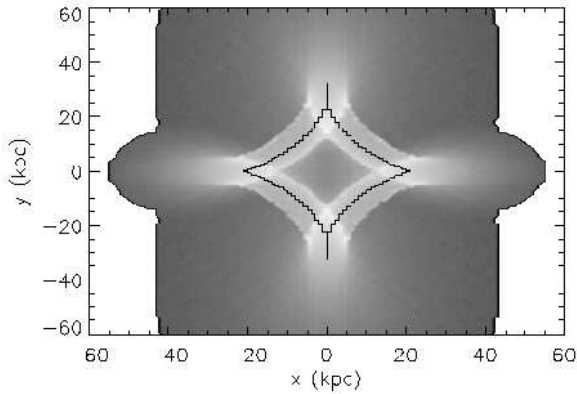}
\caption{$\lw$ map for a singular isothermal ellipsoid with
  $\sigma_v=10^3 \kms$ and $q=0.8$ aligned with the $y$ axis.  The map
  shows the function $\lw(\bm{x})$, where $\lw(\bm{x})$ is the largest
  $\lw$ value of all tangential arcs created by a source centered on
  $\bf{x}$.  $\lw$ values are low for dark regions and high for bright
  regions.  The white outside region was not searched for arcs.
  Finally, the dark, diamond shape solid line is the lens's caustic.
  There is a clear discontinuity in the $\lw$ map all around the
  caustic, corresponding to sources which touch the caustic, and whose
  longest arcs are therefore formed through the merging of two ({\em
    fold arcs}) or three ({\em cusp arcs}) distinct images. The
  dichotomy between distortion and image merging arcs is also
  reflected in the shape of the cross section curve $\sigma(\lw)$, as
  seen Fig.~\ref{fig:sie_cs}.}
\label{fig:sie_lwmap}
\end{figure}

The $\lw$ map of a cluster is defined as the function $\lw(\bm{x})$ on
the source plane, where $\lw(\bm{x})$ is the maximum length to width
ratio among all tangential arcs produced by a source centered on
$\bf{x}$.  Figure~\ref{fig:sie_lwmap} shows the $\lw$ map for an SIE
profile with velocity dispersion $\sigma_v=10^3 \kms$ and axis ratio
$q=0.8$.  The topography of the map for the lens is quite striking.
In particular, there is a discontinuity all around the lens's caustic
(shown in the figure as a solid black line). The width of this
``ribbon'' is approximately $2R$ where $R$ is the radius of the
source, indicating that all sources placed within this ribbon touch
the caustic, and therefore correspond to sources in which distinct
images merge to form a single long arc.


We therefore introduce a distinction between arcs formed through
distortion of a source, and arcs formed through the merging of several
images.  Moreover, a closer look at the ribbon around the lens's
caustic reveals that there is an additional discontinuity around the
lens's cusps, where not two but three images merge to produce a single
arc.  To differentiate between these two types of arcs, we shall refer
to them as {\em fold arcs} and {\em cusp arcs}, respectively.

\begin{figure}[t]
\epsscale{1.2}
\plotone{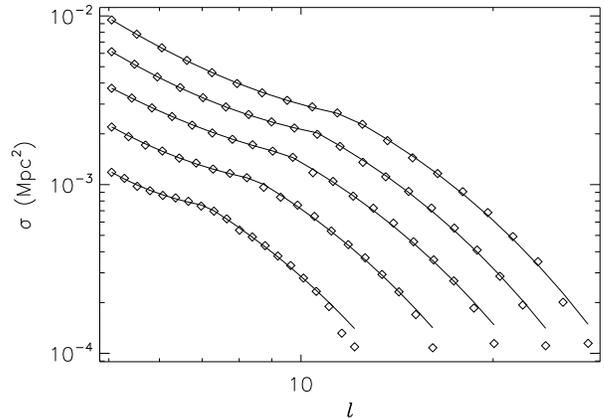}
\caption{Lensing cross section $\sigma(\lw)$ curves for SIE profiles
  with axis ratio $q=0.7$.  The velocity dispersions of the SIE halos
  are, from top to bottom, $\sigma_v/(10^3 \kms) =
  1.0,0.9,0.8,0.7,0.6$, and the assumed source radius is $5\ \kpc$.
  The cross section function exhibits a transition between a roughly
  power law regime at low $\lw$ values, and a roughly exponential
  regime at high $\lw$.  The $\lw$ value at which the transition
  occurs ($\lw_t$) corresponds to the scale at which the lensing cross
  section switches from being dominated by distortion arcs to being
  dominated by image merging arcs.  The points show the actual
  results; the solid lines show phenomenological fits discussed in
  Appendix \ref{app:fits}.}
\label{fig:sie_cs}
\end{figure}

The difference between distortion arcs and image merging arcs is
readily apparent in the shape of the cross section function
$\sigma(\lw)$, shown in Figure~\ref{fig:sie_cs} for various SIE
profiles at fixed axis ratio $q=0.7$ and for velocity dispersions
$\sigma_v/(10^3 \kms) = 0.6,0.7,0.8,0.9,$ and $1.0$.  Each cross
section curve exhibits a clear transition, or knee, at some length to
width ratio $\lw_t$ where it goes from being distortion dominated and
roughly power-law, to image merging dominated and roughly exponential
in nature.  The value of this transition scale $\lw_t$ clearly scales
with the Einstein radius of the lens, reflecting the fact that sources
will be more strongly distorted before touching the caustic for more
massive clusters.


Since the generic shape of the cross section function $\sigma(\lw)$
--- a power law at low $\lw$, and a falling exponential at high $\lw$
--- is mainly driven by the difference between distortion arcs and
image merging arcs, it is not surprising to learn that this shape is
robust to changes in the halo parameters.  Our main goal for the rest
of this section is to understand how the various features of the
lensing cross section function (its amplitude, the position of the
transition scale, etc.) depend on the various halo parameters, so that
we may better interpret the cross section functions from the
numerically simulated clusters.


\subsection{The Role of Ellipticity}
\label{sec:role_ellip}

\begin{figure}[t]
\epsscale{1.2}
\plotone{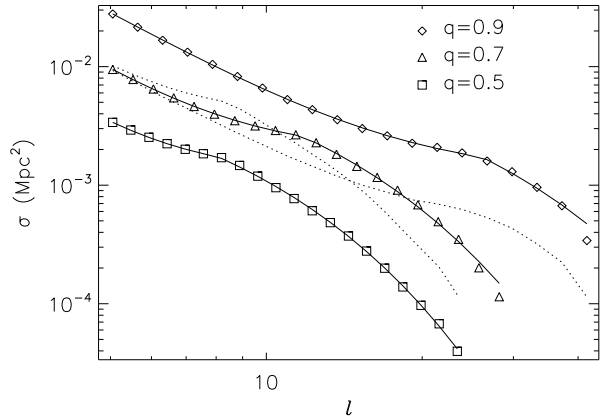}
\caption{Lensing cross sections for SIE profiles with a velocity
  dispersion $\sigma_v=10^3 \kms$ and axis ratios $q=0.5,0.7,0.9$.
  For illustration purposes, we have displaced the $q=0.9$ data
  upwards by a factor of three, and the $q=0.5$ data downwards by the
  same amount. The solid lines are our best fit models.  We see that
  the transition scale $\lw_t$ moves towards lower $\lw$ values as the
  ellipticity of the halo increases.  Also shown with dotted lines are
  all cross section curves with their natural amplitudes.  We see that
  as the halo becomes more elliptical, the amplitude of the image
  merging contribution to the cross section increases, reflecting the
  increase in the length of the tangential caustic.  The amplitude of
  the image distortion contribution, however, appears to be roughly
  independent of halo ellipticity (at fixed Einstein radius).}
\label{fig:vary_ellip}
\end{figure}

Figure~\ref{fig:vary_ellip} shows the cross section function
$\sigma(\lw)$ of three SIE profiles with velocity dispersion
$\sigma_v=10^3 \kms$ and axis ratios $q=0.9$ (diamonds), $q=0.7$
(triangles), and $q=0.5$ (squares).  For illustration purposes, we
have displaced the $q=0.9$ data upwards by a factor of three, and the
$q=0.5$ data downwards by the same amount.  The solid lines going
through the points represent power law - exponential fits (see
Appendix \ref{app:fits}).  It is evident that the transition scale
$\lw_t$ depends strongly on ellipticity.  In particular, the more
circular the profile, the larger the value for the transition scale
$\lw_t$.  This reflects the fact that for a circular profile, merging
images correspond to rings and typically have large $\lw$ ratios of
order $\lw\approx(\pi b)/(2R)$.  As the halo becomes more elliptical,
arcs straighten out and the images no longer curve around the Einstein
radius of the cluster, resulting in a smaller transition scale
$\lw_t$.


Also shown in Figure~\ref{fig:vary_ellip} as dotted lines are the
un-displaced (i.e., properly normalized) cross section curves for the
$q=0.9$ and $q=0.5$ SIE halos.  These curves demonstrate that as the
halo becomes more elliptical the \it amplitude \rm of the image
merging contribution (high $\lw$) to the cross section increases
reflecting the fact that more elliptical halos have longer tangential
caustics.  More surprising, however, is the fact that the amplitude of
the image distortion contribution (low $\lw$) appears to be roughly
independent of ellipticity, as evidenced by the fact that all three
halos have almost identical cross sections at $\lw\approx 5$. We can
understand this result qualitatively as follows: if we take a
spherical halo and deform it into an elliptical one, the shear induced
by the ellipticity will half the time add to the monopole shear, and
half the time subtract from it.  Consequently, the net cross section
will be roughly unaffected, as illustrated above.

There is one last point of interest that can be garnered from
Figure~\ref{fig:vary_ellip}: the maxim that elliptical profiles make
better lenses is not always correct.  More specifically, given a halo
of a specified Einstein radius and a $\lw$ ratio of interest, the
ellipticity with the highest cross section is that for which the
transition scale $\lw_t$ is equal to the $\lw$ ratio of interest.  For
instance, since the cross section quickly falls beyond $\lw_t$, at
high $\lw$ values ($\lw \gtrsim \pi b/2R$) and fixed Einstein radius
only circular profiles are effective lenses.


\subsection{The Role of the Density Slope}
\label{sec:sloperole}


\begin{figure}[t]
\epsscale{1.0}
\plottwo{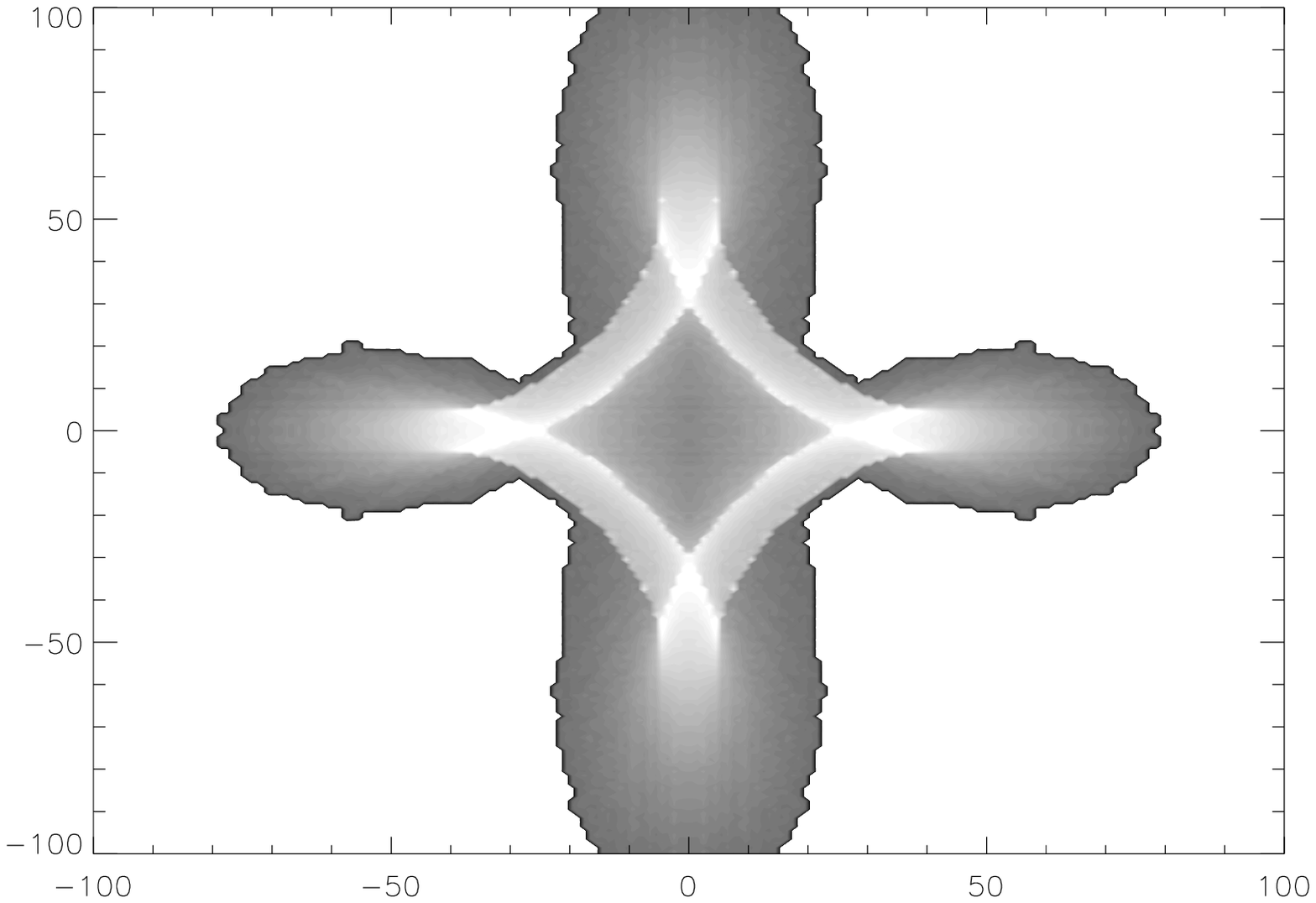}{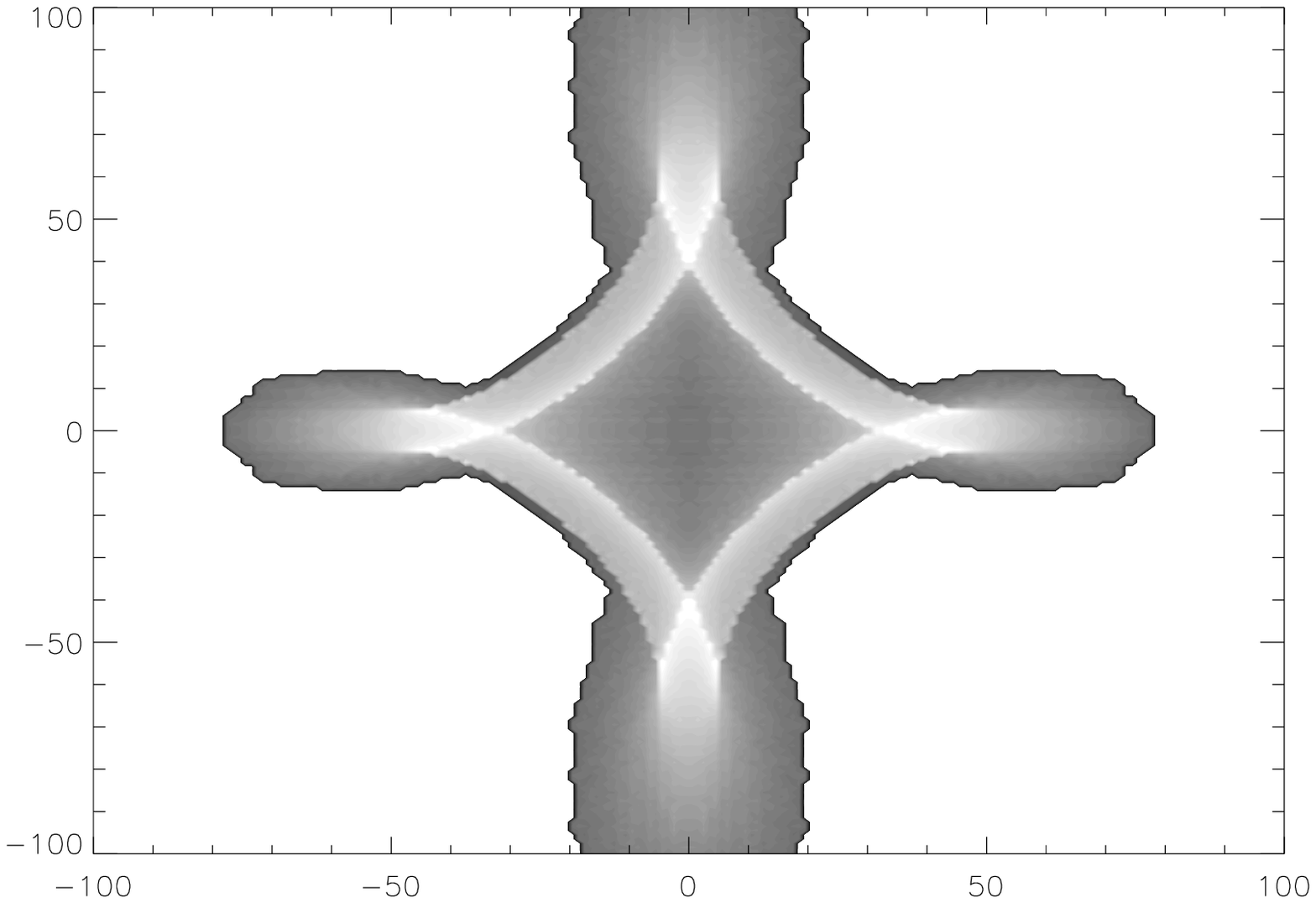}
\plottwo{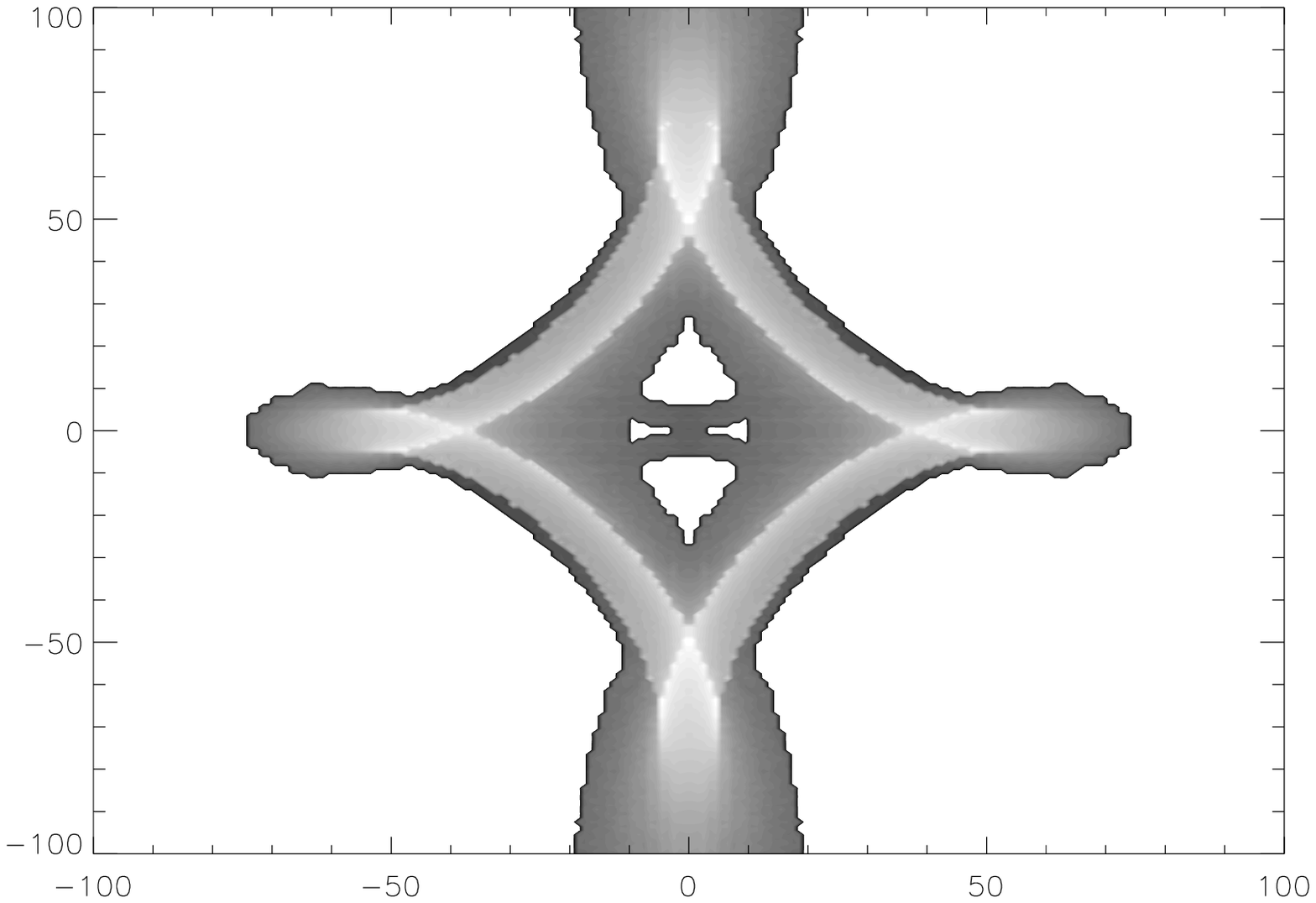}{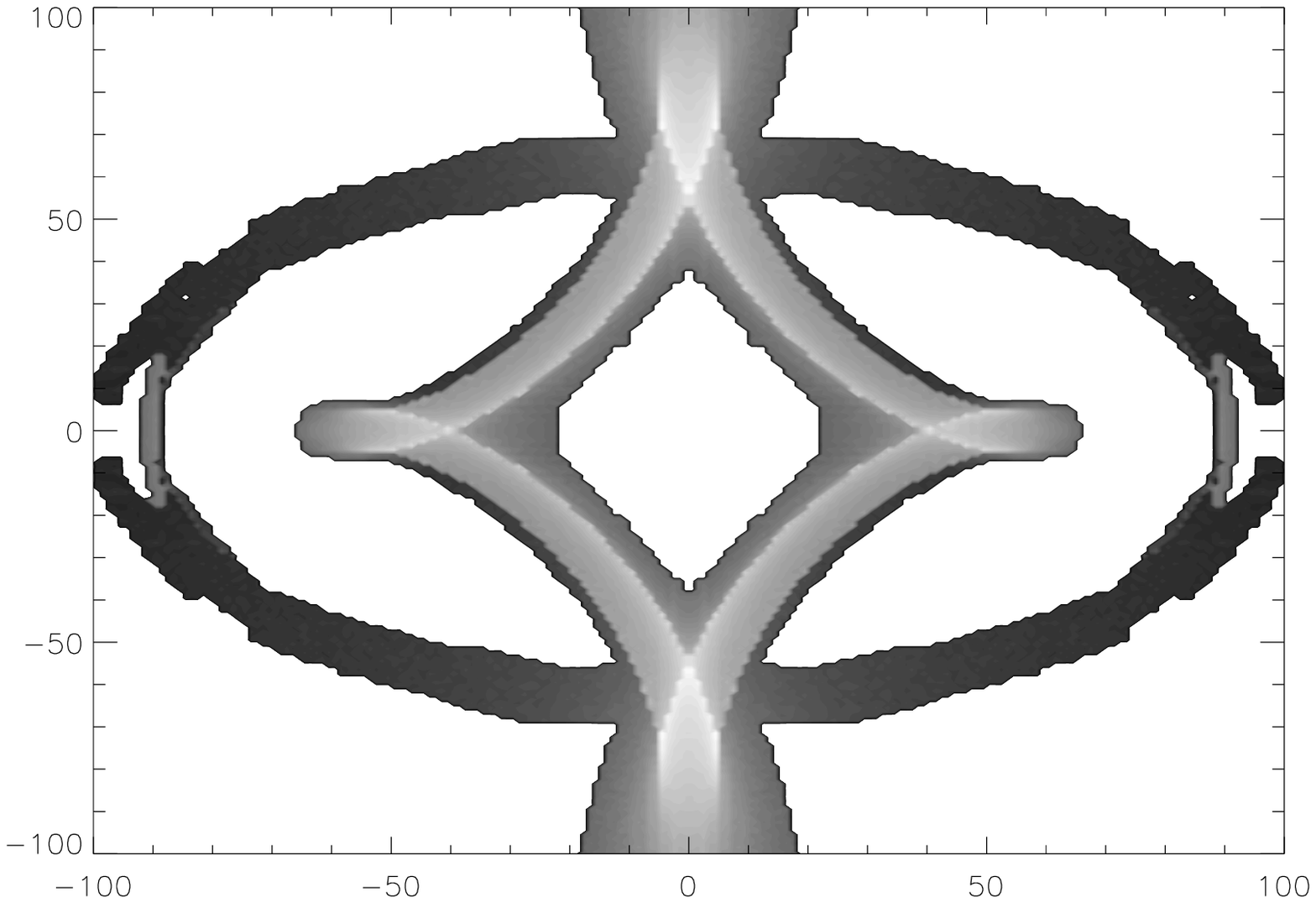}
\plottwo{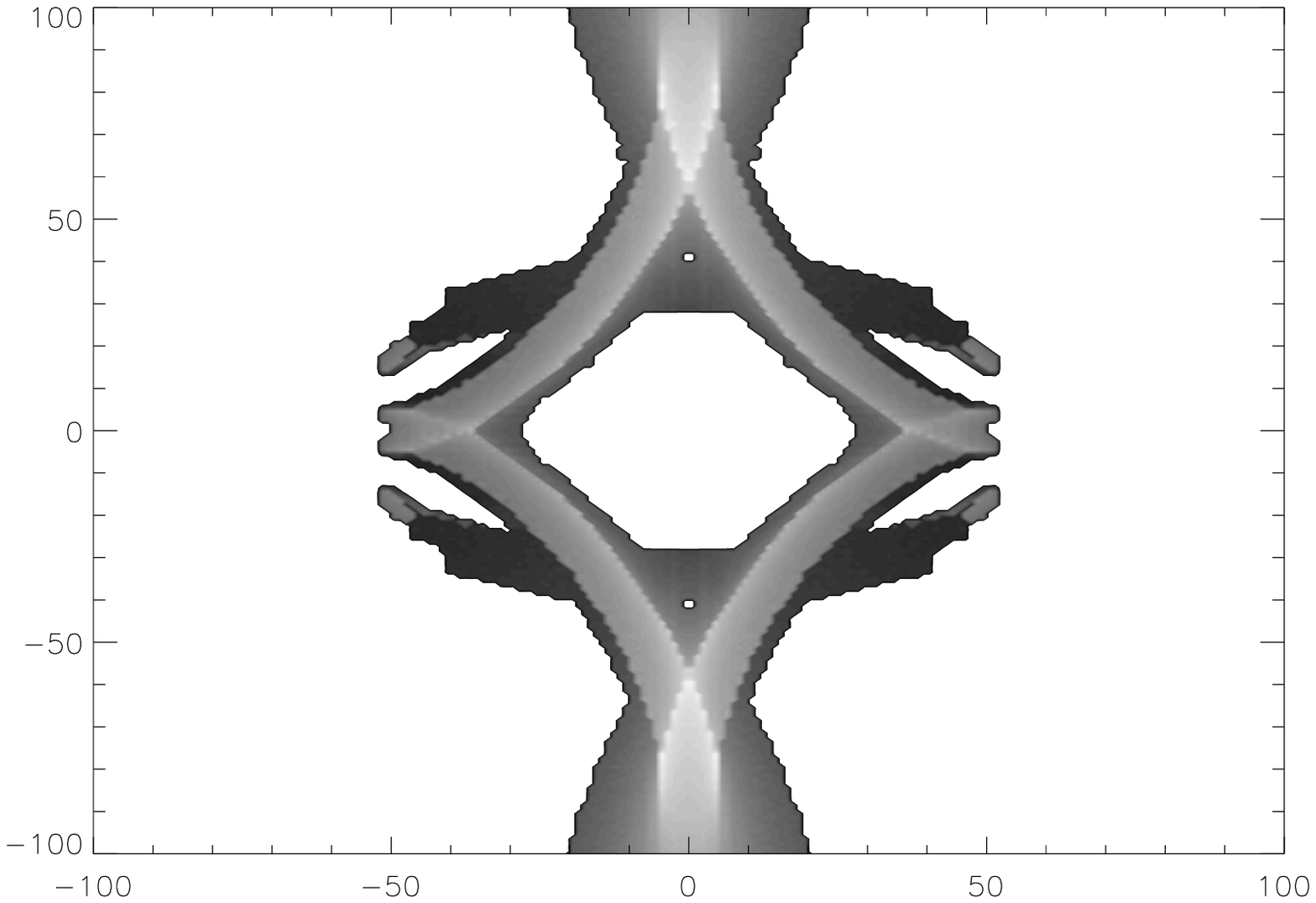}{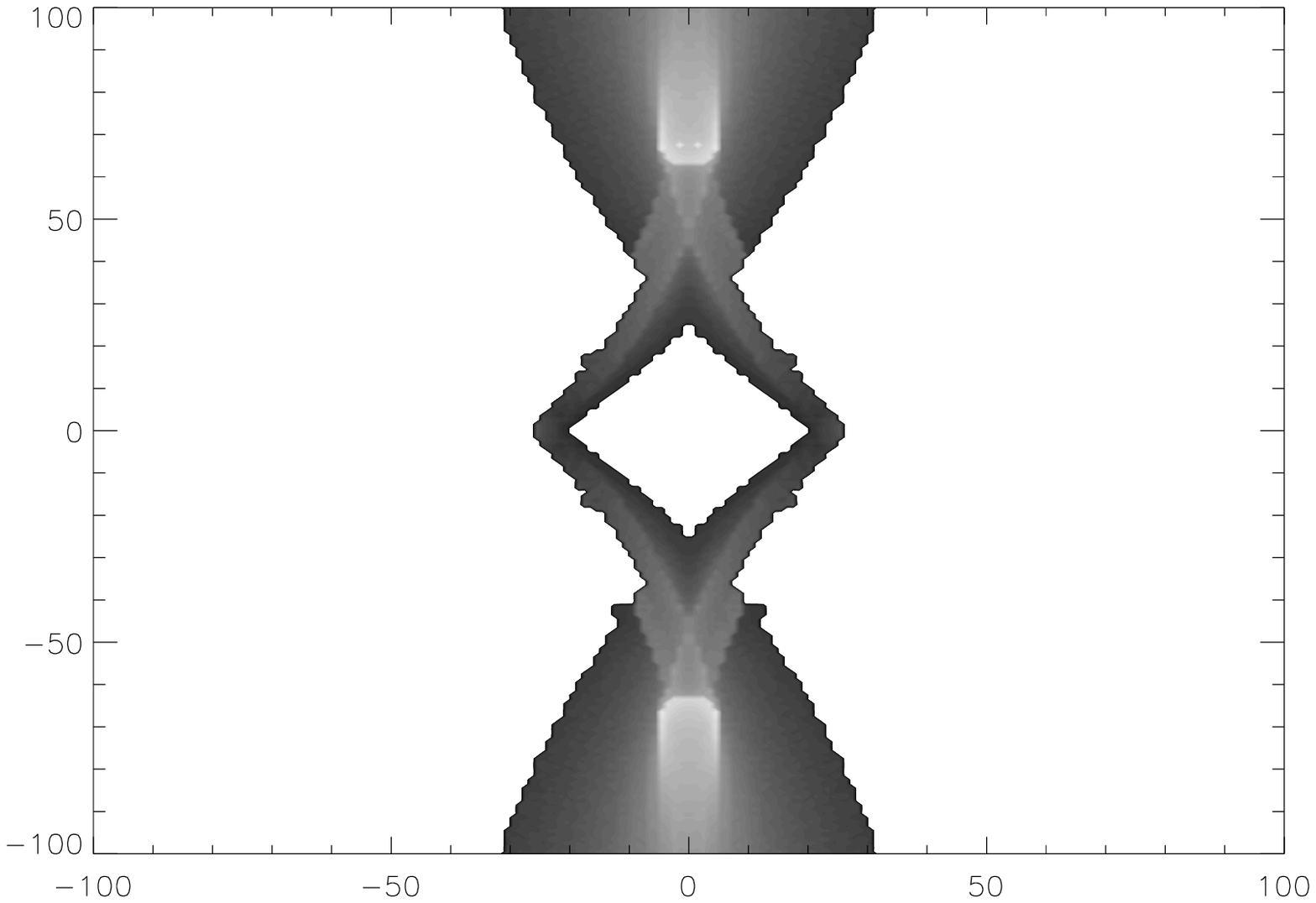}
\caption{$\lw$-maps for power-law profiles with axis ratio $q=0.6$,
  for profile slopes $\beta=1.4$ (top-left), $\beta=1.2$ (top-right),
  $\beta=1.0$ (middle-left), $\beta=0.8$ (middle-right), $\beta=0.6$
  (bottom-left), and $\beta=0.4$ (bottom-right).  The long axis of the
  lens is always along the $y$ axis.  The Einstein radius is fixed to
  that of an SIS with a velocity dispersion $\sigma_v=10^3\ \kms$.  We
  have used the same gray scale in all maps to allow easy comparison
  between them.  Note that fold arcs and distortion arcs not aligned
  with the long axis of the lens quickly become smaller (have smaller
  $\lw$ values) as $\beta$ decreases.  For extremely shallow profiles,
  this reduction in $\lw$ can be so severe that fold arcs may become
  much smaller than single image arcs along the long axis of the lens,
  and the fold arc contribution to the cross section becomes
  unresolved.  As a final note, the dark rings seen in some of the
  maps are due to sources placed along the radial caustic after we
  filter out the $\lw$ values from the corresponding radial arcs.}
\label{fig:e0.4maps}
\end{figure} 


To understand how the slope of the density profile affects lensing, we
first look at the $\lw$ maps of a series of analytic models.
Figure~\ref{fig:e0.4maps} shows the maps for various power-law
profiles where we fixed the Einstein radius to that of an SIS with
velocity dispersion $\sigma_v=10^3 \kms$ and the axis ratio to
$q=0.6$.  In order to facilitate comparison between the various maps,
we have used the same gray scale in all panels.

We see that the division of arcs into single image, fold, and cusp
arcs is generic.  What is striking is the strong impact of the profile
slope on the length to width ratios of fold, and to a lesser extent,
cusp arcs. The ``ribbon'' around the lens's caustic decreases
monotonically in intensity as $\beta$ decreases, and is even truncated
in the case of the $\beta=0.4$ profile.  The cusp arcs of the flatter
profiles also appear to have smaller length to width ratios than their
counterparts in steep profiles, but the difference is less extreme.

A similar pattern is also evident in the image distortion component of
the maps, where we see that the length to width ratio of sources not
aligned with the long axis of the lens decreases quickly as the
profile becomes shallower.  We can explain these patterns through a
heuristic argument as follows.  Consider a small source located off
the long axis of the lens.  If the halo is highly elliptical, the
local curvature of the mass distribution is low and hence the shear
induced through ellipticity is low.  Likewise, flattening of the
profile reduces the shear induced by the mass distribution by making
it more uniform, so flat, elliptical profiles are poor lenses if the
source is off the long axis of the lens.  If a source is along the
long axis of the lens, however, the curvature of the mass distribution
is high and the resulting images are strongly distorted, even if the
profile itself is shallow.  Thus, highly elliptical flat profiles will
only produce arcs when the sources are placed along the long axis of
the lens.


\begin{figure}[t]
\epsscale{1.2}
\plotone{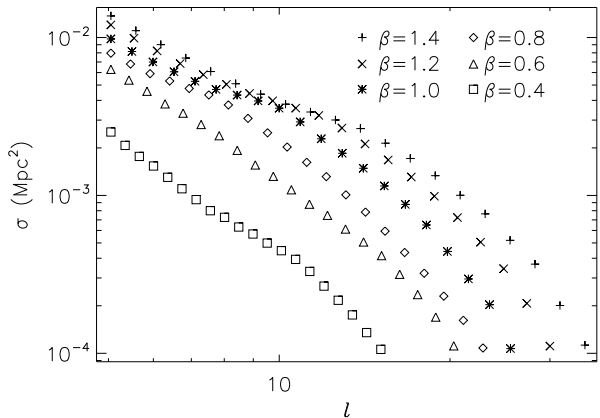}
\caption{Cross sections for power law profiles of various slopes
  $\beta$.  The axis ratio is fixed at $q=0.6$, and the Einstein
  radius is that of an SIS with a velocity dispersion $\sigma_v=10^3
  \kms$.  The corresponding $\lw$ maps can be seen in
  Fig.~\ref{fig:e0.4maps}.  Note that the curves for $\beta\geq 0.8$
  are all similar in shape, with the transition scale $\lw_t$
  decreasing monotonically with $\beta$, reflecting the lower $\lw$
  value of fold arcs for shallow profiles.  At $\beta=0.6$ the profile
  becomes so flat that single image arcs along the long axis of the
  lens can be substantially larger than fold arcs.  Consequently, the
  single image contribution to the cross section ``spills over'' into
  the fold arc regime, qualitatively changing the features of the
  cross section.  Further flattening of the profile --- the
  $\beta=0.4$ case --- results in such small fold arcs that their
  contribution to the cross section becomes completely washed out by
  the single image contribution.  The cusp arc contribution, however,
  becomes prominent, and the generic shape of the cross section is
  once again similar to that of the steeper $\beta\gtrsim 0.8$
  profiles.}
\label{fig:slope}
\end{figure} 


Let us now turn our attention to the cross section function
$\sigma(\lw)$, shown in Figure~\ref{fig:slope} for the sample cases
from Figure~\ref{fig:e0.4maps}.  Focus first on the top four curves,
corresponding to $\beta=1.4,1.2,1.0,$ and $0.8$.  All four of these
have the characteristic two-component shape, with the transition scale
$\lw_t$ decreasing as the profile becomes shallower.  This agrees with
our expectations: the transition scale $\lw_t$ for these curves is
roughly the minimum length to width ratio of fold arcs.  Since fold
arcs become smaller as the profile becomes shallower, the
corresponding transition scale decreases.

Turning to the $\beta=0.6$ curve, however, we see that there is no
obvious transition.  Furthermore, the $\beta=0.4$ curve has a
transition that occurs at larger values than the transition for the
$\beta=0.8$ case.  This behavior may seem puzzling at first, but is
easily understood as follows.  As $\beta$ decreases from $\beta=0.8$,
fold arcs keep getting smaller and smaller, so at some point
distortion arcs from sources placed along the long axis of the lens
become longer than off-axis fold arcs.  In other words, the distortion
component of the cross section will ``spill over'' into the fold arc
region, thereby erasing the clear transition observed in the steeper
profiles.  As $\beta$ is further decreased, fold arcs become
negligible, and a new transition becomes evident, but now the
transition is between distortion arcs and cusp arcs.  This explains
why the $\beta=0.4$ transition occurs at higher length to width ratios
than the $\beta=0.8$ transition.

There is an additional trend apparent in the cross section functions
shown in Figure~\ref{fig:slope}: as the profile gets shallower, the
amplitude of the image distortion contribution to the cross section
decreases with $\beta$, even though the Einstein radius of the various
profiles has been held fixed. This reflects the fact that flat
profiles induce little gravitational shear, and are thus not very
effective at distorting the images of sources.


\section{Simulated Clusters}
\label{sec:simulations}


\subsection{The Cosmological Simulations}

Next, we analyze high-resolution cosmological simulations of six
cluster-size systems in the ``concordance'' flat {$\Lambda$}CDM model:
$\Omega_M = 1-\Omega_{\Lambda}=0.3$, $\Omega_b = 0.04286$, $h=0.7$ and
$\sigma_8=0.9$, where the Hubble constant is defined as $H_0 = 100\,h
\kms \Mpc^{-1}$, and $\sigma_8$ is the power spectrum normalization on
an $8h^{-1}$~Mpc scale. The simulations were done with the Adaptive
Refinement Tree (ART) $N$-body$+$gasdynamics code \citep{kravtsov99,
  kravtsov_etal02}, a Eulerian code that uses adaptive refinement in
space and time, and (non-adaptive) refinement in mass
\citep{klypin_etal01} to reach the high dynamic range required to
resolve cores of halos formed in self-consistent cosmological
simulations. The simulations presented here are a subset of the
simulated cluster sample presented in \citet{kravtsov_etal06} and
\citet{nagai_etal06b}, and we refer the reader to these papers for
more details.  Here we summarize the main parameters of the
simulations and list the basic properties of clusters at $z$=0 in
Table~\ref{tab:sim}.

High-resolution simulations were run using a 128$^3$ uniform grid and
8 levels of mesh refinement in the computational boxes of
$120h^{-1}$~Mpc for CL1-CL3 and $80h^{-1}$~Mpc for CL4-6, which
corresponds to the dynamic range of $128\times 2^8=32768$ and peak
formal resolution of $80/32768 \approx 2.44h^{-1}$~kpc, corresponding
to the actual resolution of $\approx 2\times 2.44\approx 5h^{-1}$~kpc.
Only the region of $\sim$3--10$h^{-1}$~Mpc around the cluster was
adaptively refined, while the rest of the volume was followed on a
uniform $128^3$ grid. The mass resolution corresponds to the effective
$512^3$ particles in the entire box, or the Nyquist wavelength of
$\lambda_{\rm Ny} = 0.469h^{-1}$ and $0.312h^{-1}$ {\em comoving} Mpc
for CL1-3 and CL4-6, respectively, or $0.018h^{-1}$ and $0.006h^{-1}$
Mpc in the physical units at the initial redshift of the simulations.
The dark matter particle mass in the region around the cluster was
$9.1\times 10^{8} h^{-1}\ \msun$ for CL1-3 and $2.7\times 10^{8}
h^{-1}\ \msun$ for CL4-6, while other regions were simulated with
lower mass resolution.

We repeated each cluster simulation with and without radiative cooling
and processes of galaxy formation.  The first set of ``adiabatic''
simulations have included only the standard gas dynamics for the
baryonic component without gas cooling and star formation.  The second
set of simulations included gas dynamics and several physical
processes critical to various aspects of galaxy formation: star
formation, metal enrichment and thermal feedback due to supernovae
(type II and type Ia), self-consistent advection of metals,
metallicity dependent radiative cooling and UV heating due to
cosmological ionizing background \citep{haardt_madau96}. Throughout
this paper, we refer to the adiabatic simulations and simulations with
cooling and star formation simply as ``adiabatic'' and ``cooling''
run, respectively.

In the cooling run, the cooling and heating rates take into account
Compton heating and cooling of plasma, UV heating, and atomic and
molecular cooling and are tabulated for the temperature range
$10^2<T<10^9$~K and a grid of metallicities, and UV intensities using
the {\tt Cloudy} code \citep[ver.  96b4,][]{ferland_etal98}. The
Cloudy cooling and heating rates take into account metallicity of the
gas, which is calculated self-consistently in the simulation, so that
the local cooling rates depend on the local metallicity of the gas.
Star formation in these simulations was done using the
observationally-motivated recipe \citep[e.g.,][]{kennicutt98}:
$\dot{\rho}_{\ast}=\rho_{\rm gas}^{1.5}/t_{\ast}$, with
$t_{\ast}=4\times 10^9$~yrs. The code also accounts for the stellar
feedback on the surrounding gas, including injection of energy and
heavy elements (metals) via stellar winds and supernovae and secular
mass loss.

Starting from the well-defined cosmological initial conditions, these
simulations follow the formation of galaxy clusters and capture the
dynamics of dark matter, stars and ICM self-consistently.  These
simulations can therefore be used to examine the effects of gas
cooling and star formation on the density profiles
\citep[e.g.,][]{gnedinetal04} and shapes \citep{kazantzidisetal04} of
dark matter halos as well as the roles of substructure present in a
realistic cosmological context. Although the magnitude of the effects
are likely overestimated in the current simulations due to overcooling
problem, the effects of baryon cooling we discuss below are generic.





\begin{deluxetable}{lcccc}
\tablecaption{Properties of simulated clusters from cooling runs at z=0}
\tablehead{
\\
\multicolumn{1}{c}{Name} &
\multicolumn{1}{c}{}&
\multicolumn{1}{c}{$R_{\rm 500c}$} &
\multicolumn{1}{c}{$M_{\rm 500c}^{\rm gas}$}  &
\multicolumn{1}{c}{$M_{\rm 500c}^{\rm tot}$} 
\\ \\
\multicolumn{2}{c}{} &
\multicolumn{1}{c}{($h^{-1}$~Mpc)} &
\multicolumn{1}{c}{($10^{13} h^{-1}\ \msun$)} & 
\multicolumn{1}{c}{($10^{14} h^{-1}\ \msun$)}
\\
}
\startdata
\\
CL1  & & 1.160 & 8.19 & 9.08 \\
CL2  & & 0.976 & 5.17 & 5.39 \\
CL3  & & 0.711 & 1.92 & 2.09 \\
CL4  & & 0.609 & 1.06 & 1.31 \\
CL5  & & 0.661 & 1.38 & 1.68 \\
CL6  & & 0.624 & 1.22 & 1.41 \\
\enddata
\label{tab:sim}
\end{deluxetable}
\vspace{0.8cm}


\subsection{Surface Density Maps and Ray Tracing}

To create the surface density map of a cluster, we begin by
nesting $2 \hMpc$ and $8 \hMpc$ boxes on the center of the
cluster (defined as the position of the most bound dark matter
particle in the halo).  The boxes have $256$ and $64$ pixels
to a side, respectively, and the density at each grid point is
computed with a cloud in cell algorithm.  The 3D density maps
are projected along the region $0\le r/(1 \hMpc) \le 1$ and
$1 \le r/(1 \hMpc) \le 4$ for the small and large boxes
respectively, and the resulting surface density maps added to
produce a 2D surface density map which is $8 \hMpc$ in extent
and reaches a resolution of $7.8 \hkpc$ in the central regions.

Given the surface density map, we use Fast Fourier Transforms to
compute the angular deflection and inverse magnification tensor at
each lens plane grid-point.  We then select a small region near the
center of the lens plane,\footnote{Operationally, a small square
section of the lens plane is selected by demanding that no pixel
within $50\ \kpc$ of the edge of the square have an eigenvalue ratio
$|\lr/\lt|>0.9\lwmin$.  This ensures all pixels relevant for
lensing are well within the selected region of space.} and refine the
lens plane grid using bilinear interpolation.  The refined lens plane
resolution $r_l$ is defined in terms of the source plane resolution
$r_s$ via $r_l = r_s/\mbox{max}(\lr)$ where $r_s$ is the source plane
grid resolution and max($\lr$) is the maximum value of the radial
eigenvalue $\lr$ over all pixels where $|\lr/\lt|>0.9\lwmin$
(i.e., all pixels relevant for lensing).  The source plane resolution
$r_s$ is itself defined via $r_s = R/N$ where $R$ is the source radius
and $N$ is the number of pixels that ``fit'' within $R$, for which we
take $N=5\sqrt{2}$ as our default value.  Finally, once the lens and
source plane grids are defined, we link each source plane pixel to
every lens plane pixel that maps to it, resulting in a lookup table
that can quickly generate the lensed image of any source on the source
grid.


\subsection{The Impact of Gas Cooling in Lensing Cross Sections}
\label{sec:isolated}


\begin{figure}[t]
\epsscale{1.2}
\plotone{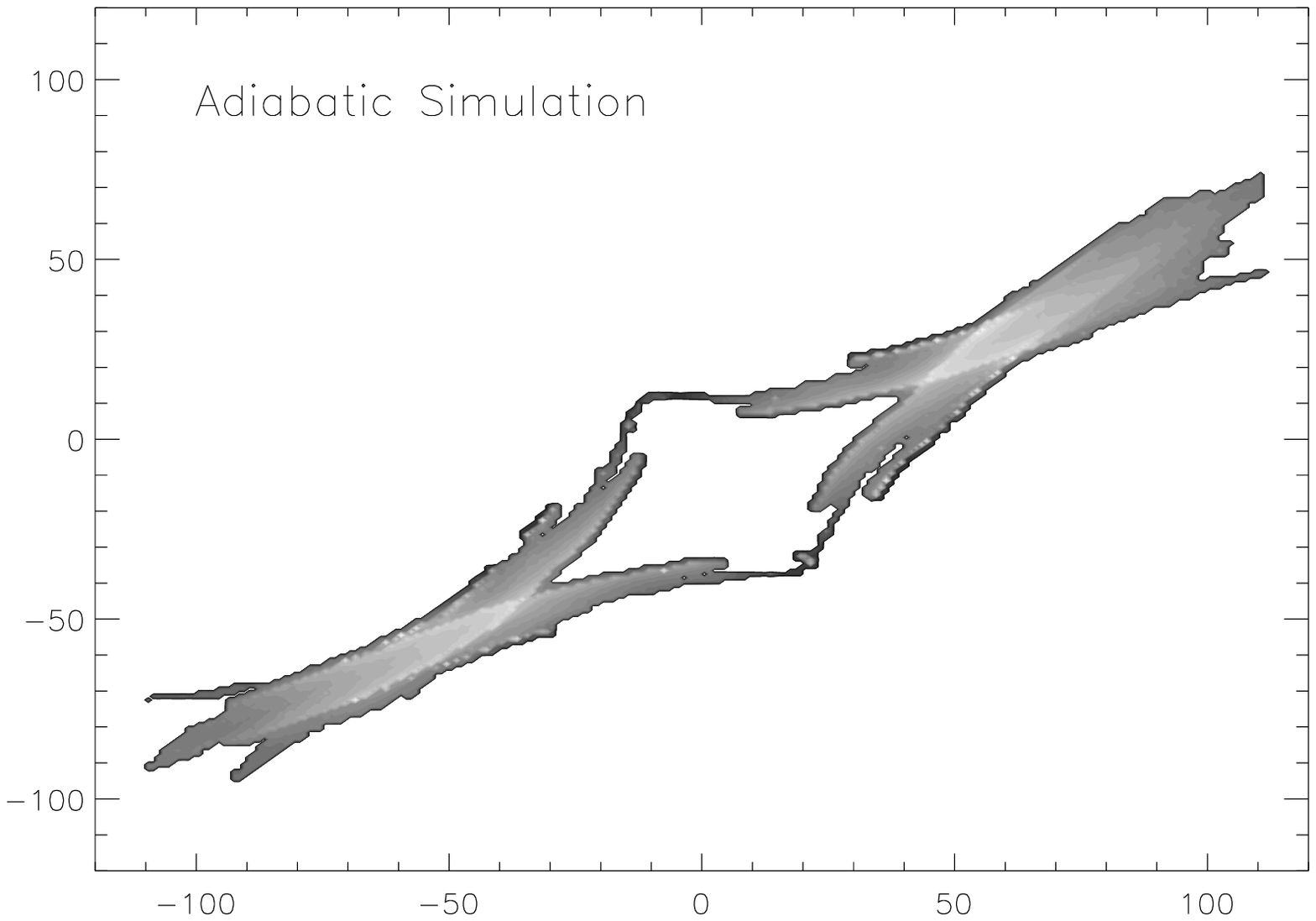}
\plotone{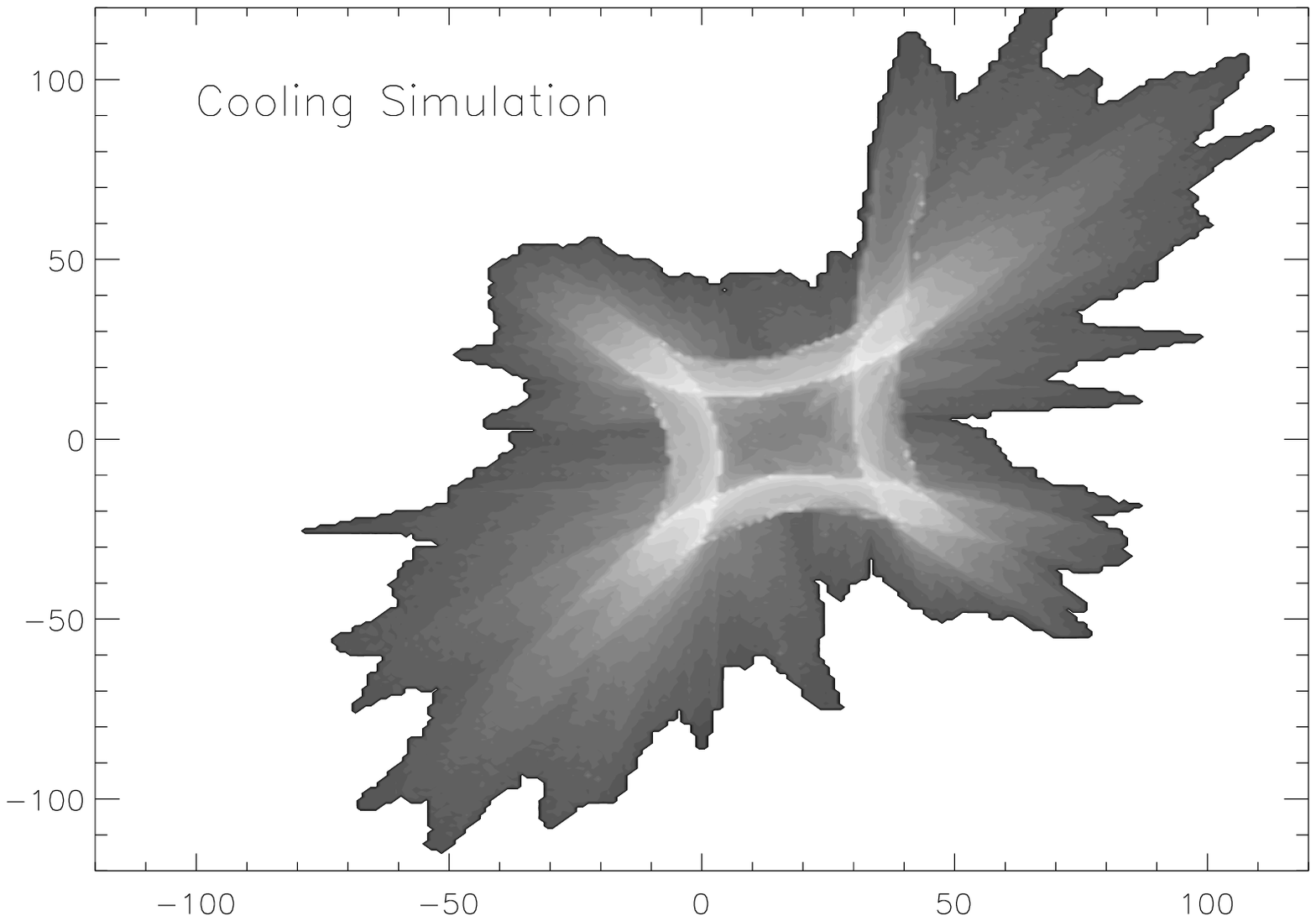}
\caption{$\lw$ maps for CL2 in the adiabatic (top) and
cooling (bottom) simulations.  We have used the same gray scale
for both maps to facilitate comparison.  The difference between
the two maps is dramatic.  In particular, adiabatic clusters can
produce arcs only for sources located along the major axis of the
lens, and these are cusp arcs; fold arcs provide a negligible
contribution to the lensing cross section.  Cooling clusters,
on the other hand, exhibit a much more isotropic arc distribution,
and should exhibit prominent fold arcs.  These qualitative
features are exactly what we would expect given that adiabatic
clusters tend to be flat and highly elliptical, whereas cooling
clusters are less elliptical and have much steeper profiles in
the inner regions of the cluster.  These maps are characteristic
of all adiabatic and cooling maps we examined in which the central
density peak was relatively isolated.}
\label{fig:isolated_maps}
\end{figure} 


We start by analyzing one of the relaxed clusters with an
isolated central density peak to illustrate the general features.
Figure~\ref{fig:isolated_maps} shows the $\lw$ maps of the adiabatic
and cooling runs of CL2, viewed along a given line of sight.  The
difference between the two maps is dramatic: the adiabatic cluster
is strongly asymmetrical, and arcs can only be produced if sources
are placed along the long axis of the lens.  The cooling cluster,
by contrast, can produce arcs for sources at all azimuthal angles.
The cross section for fold arcs is very small in the adiabatic
cluster, but prominent in the cooling cluster.  These are exactly
the trends we would expect based on our analytic models given
that adiabatic clusters are flat and highly elliptical, whereas
cooling clusters tend to be less elliptical and have much steeper
inner density profiles.



\begin{figure}[t]
\epsscale{1.2}
\plotone{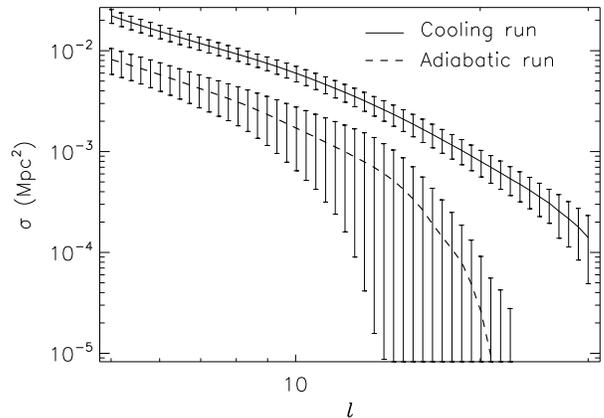}
\caption{The average cross section and its variance among 25 random
  lines of sight for a massive cluster with a relatively isolated
  central density peak in cooling (solid line) and adiabatic (dashed
  line) simulations.  Cooling boosts the average cross section by a
  factor of $\approx 4$ at low $\lw$ and significantly more at higher
  $\lw$.  In addition, it significantly reduces the variance among
  lines of sight.  Finally, note that the cooling cluster is a much
  more effective lens, capable of producing significantly larger arcs
  than the adiabatic cluster.  This implies that cooling decreases the
  lower mass cut in lensing-selected clusters.}
\label{fig:ad_sf_isolated}
\end{figure}

Next, we study the effect of gas cooling and star formation on the
cross section function $\sigma(\lw)$.  Figure~\ref{fig:ad_sf_isolated}
shows the average cross section for both the adiabatic and cooling
simulations.  The impact of baryonic cooling is immediately evident:
at $\lw=5$, the boost in the cross section due to cooling is a
factor of $\approx 4$.  The cooling cluster is capable of forming
much larger arcs than the adiabatic cluster because gas cooling
enhances the mass in the central regions of the cluster, steepens
its density profile, and makes the halo less elliptical.  All of
these effects make the transition scale $\lw_t$ shift to larger
values.  The net effect is a dramatic increase in the length to
width ratio of the arcs that the cluster can produce.  Turning
this around, we infer that the minimum mass cluster that can
produce arcs above some given $\lw$ is {\em smaller} with cooling
than without.  Reducing the effective mass cut may significantly
increase the number of lensing clusters, depending on the steepness
of the halo mass function.  We make an estimate of this effect in
section \ref{sec:abundances}.


Also striking in Figure~\ref{fig:ad_sf_isolated} is the difference in
the variance in the cross section among different lines of sight: the
cooling simulation exhibits considerably smaller variance than the
adiabatic one.  In the absence of gas cooling, the lensing cross
section varies strongly with viewing angle \citep[see
also][]{dalalholderhennawi04}. This is due to the combined effects of
the flatness of the NFW profile and triaxiality of a halo.  That is,
if the inner density profile is shallow, a small boost to the density
due to changes in the viewing angle can dramatically increase the
Einstein radius and hence the lensing cross section, resulting in a
large dispersion in the cross sections for different lines of sight.
In the cooling simulations, by contrast, the Einstein radius and cross
section of the lens remain relatively unchanged as we vary the viewing
angle.  The main reason is that gas cooling steepens the inner density
profile of the cluster, making the cross section less sensitive to
boosts in the projected mass.\footnote{Recall that at fixed slope and
  Einstein radius, the amplitude of the distortion component of the
  lensing cross section is independent of ellipticity (see section
  \ref{sec:role_ellip}).}  The fact that the cooling clusters are less
triaxial also contributes to the reduced variance, but is not a
dominant effect.


\subsection{The Impact of Substructures}
\label{sec:substructure}


\begin{figure}[t]
\epsscale{1.2}
\plotone{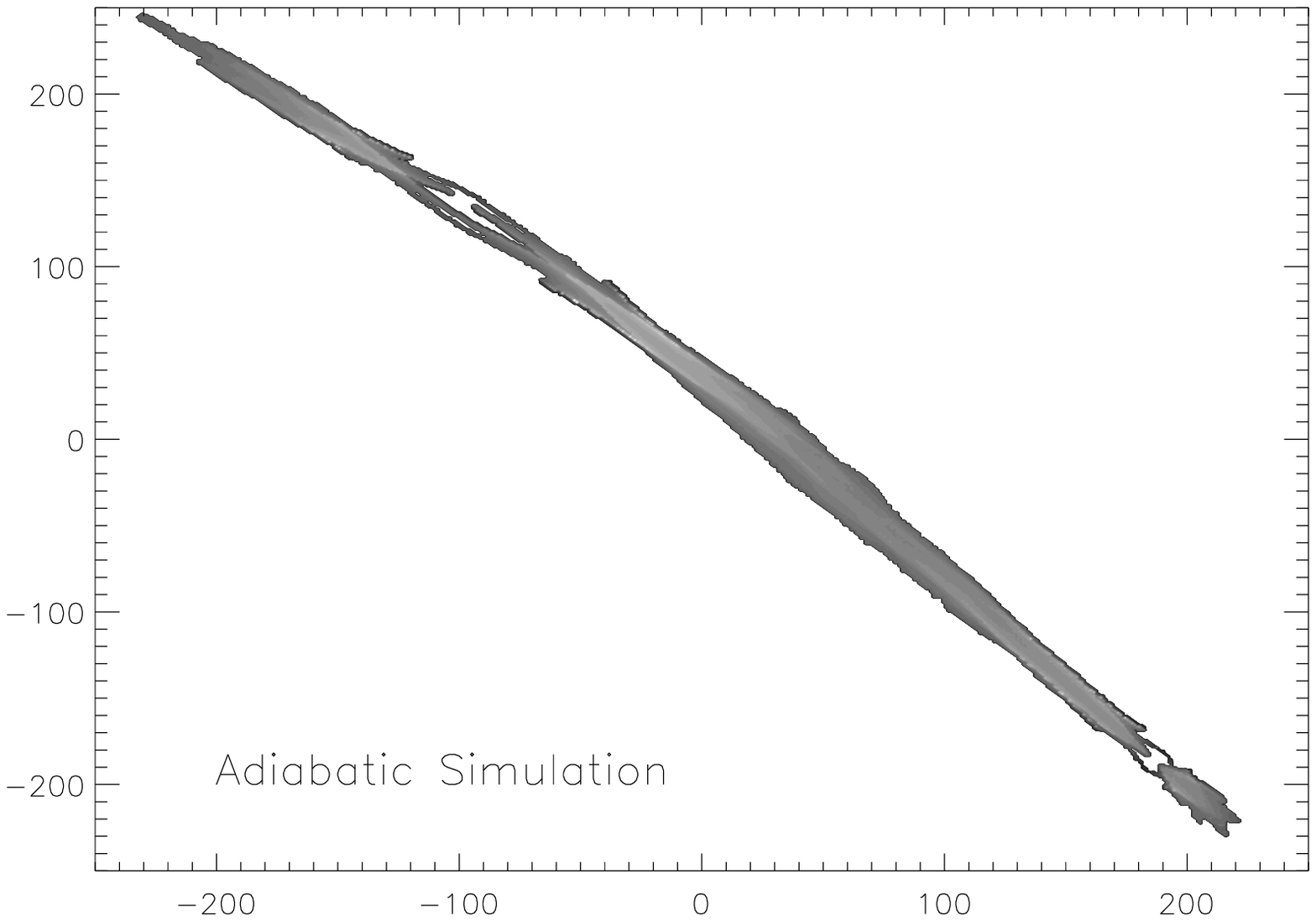}
\plotone{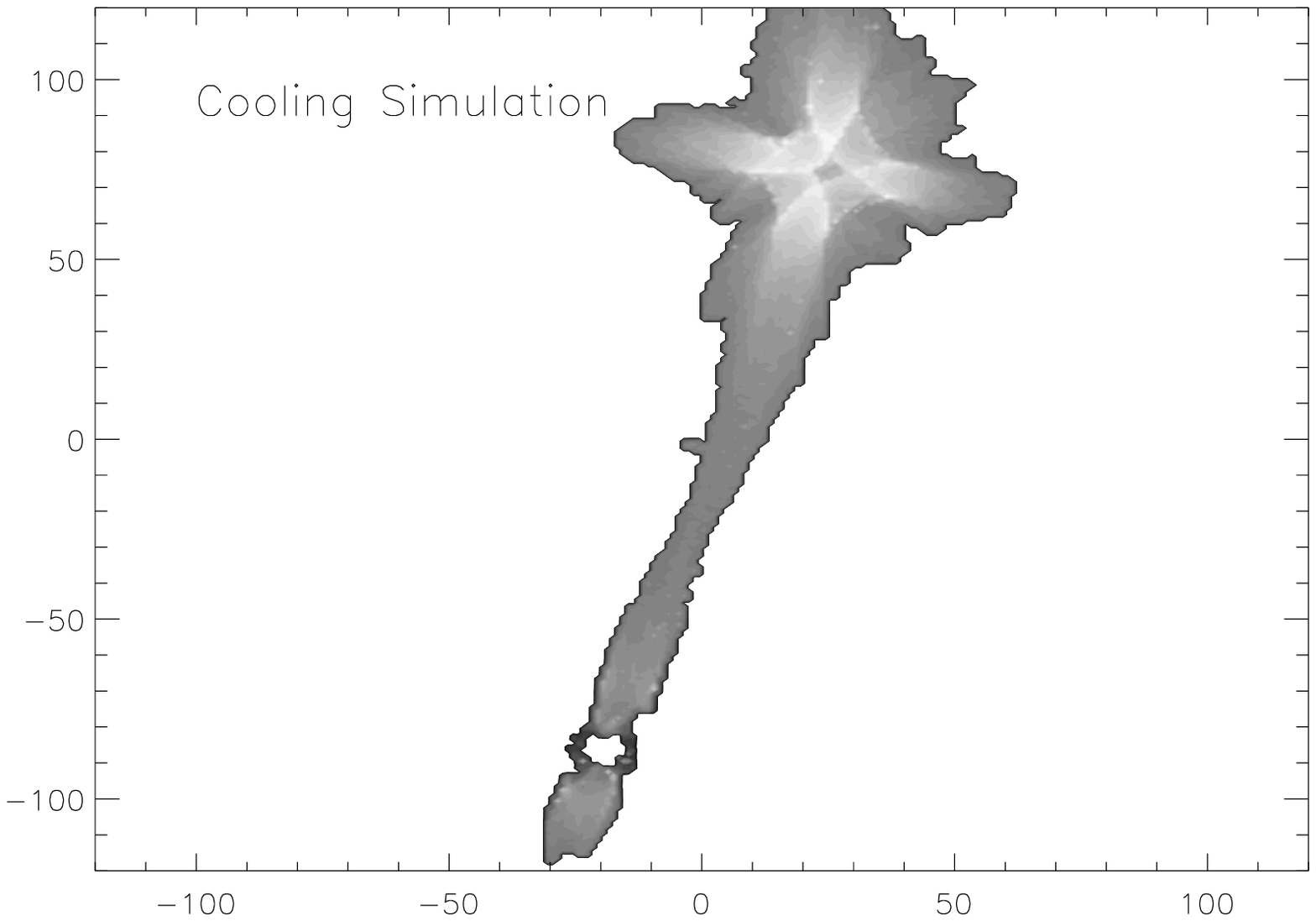}
\caption{$\lw$ maps of clusters in which there is substructure near
  the main density peak.  The top panel shows an adiabatic cluster,
  and the bottom panel shows a cooling cluster (not the same cluster).
  The basic map topology around the central density peak is largely
  unaffected by the substructure, though we often found a ``bridge''
  connecting the central density peak with the nearest substructure.
  This bridge enhances the lensing cross section for moderate length
  to width ratios, but in general we expect the effect to be modest.}
\label{fig:subs_maps}
\end{figure} 


We now examine the effects of substructure on the lensing cross
section.  Figure~\ref{fig:subs_maps} shows two typical $\ell$ maps for
clusters with substructure, for both adiabatic (top panel) and cooling
(bottom panel) simulations.  In both cases, substructure leads to the
appearance of ``bridges'' between the central density peak and the
most nearby substructure.

It is apparent from the maps, however, that such bridges provide only
a modest contribution to the lensing cross section of the largest arcs
in both the adiabatic and the cooling clusters, in qualitative
agreement with previous estimates of the impact of substructure on the
lensing cross section
\citep[][]{meneghettietal00,floresetal00,hennawietal05}.


\begin{figure}[t]
\epsscale{1.2}
\plotone{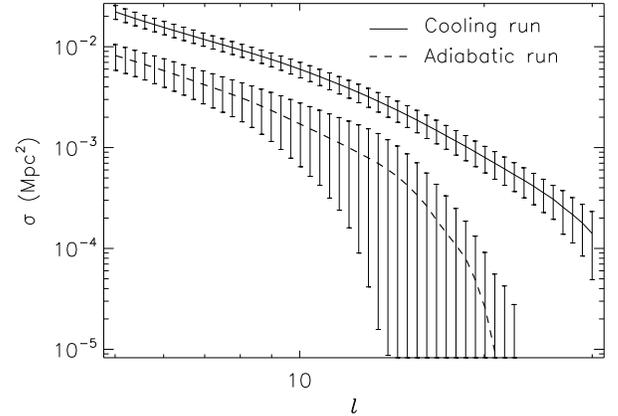}
\caption{The average cross section and its variance for our most
  massive cluster (CL1).  This cluster has substructures present near
  its central density peak, yet the average cross section function
  $\sigma(\lw)$ exhibits the same general features seen for the
  isolated central density peak case examined in
  Fig.~\ref{fig:ad_sf_isolated}.  We found this to be true of all the
  clusters we considered.  Thus, the presence of substructures does
  not appear to significantly affect the impact of baryonic cooling on
  arc abundances.}
\label{fig:subs_cs}
\end{figure}


Figure~\ref{fig:subs_cs} shows the average cross section for cluster
CL1 over 25 random line of sight projections.  Even though the central
density peak of CL1 is not isolated, the impact of baryonic cooling on
the lensing cross section is strikingly similar to that of CL2, which
did have an isolated central density peaks (cf.\
Fig.~\ref{fig:ad_sf_isolated}).  In particular, the lensing cross
section is still boosted by a factor of $\approx 3$ in the cooling
run, and the transition length to width ratio $\lw_t$ is again shifted
to larger values.  We found this to be the case in all of our
simulated clusters.  We conclude that substructures do not
significantly change how baryonic cooling affects arc formation.  A
complete quantitative characterization of these effect, however,
requires a larger cluster sample than what is available to us at this
time.

\section{The Impact of Baryonic Cooling on Giant Arc Abundance Estimates}
\label{sec:abundances}

As discussed in section \ref{sec:cs}, estimating the number of giant
arcs per unit area in the sky involves characterizing the average
lensing cross section $\avg{\sigma|m;z_l,z_s,\bp}$ for halos of mass
$m$ as a function of redshift and for a range of source structure
parameters (e.g., radius and ellipticity).  Since our lens sample is
limited to a handful of clusters, we cannot provide a robust estimate
of $\avg{\sigma|m;z_l,z_s,\bp}$, even ignoring dependencies on source
structure parameters.  It is thus evident that extrapolation of our
results is necessary, and will involve some guesswork.  In light of
these difficulties, we have opted to use the simplest possible scaling
arguments to estimate the {\em ratio} of the number of giant arcs
expected in cooling simulations to the number expected in simulations
with only adiabatic gas physics.

Our analysis begins by noting that eq.~(\ref{eq:arcab}) can be written
as a product,
\begin{equation}
\frac{dn_{arcs}}{dz_sdz_ld\bp} = g(z_s,z_l,\bp)\ \tau(\lw;z_s,z_l,\bp),
\end{equation}
where
\begin{equation}
g(z_s,z_l,\bp) = \frac{dn_s(\bp;z_s)}{d\bp}\frac{dV}{dz_s}\frac{dV}{dz_l},
\end{equation}
and
\begin{equation}
\tau (\lw;z_s,z_l,\bp) = \int dm\ \frac{dn}{dm} \sigma(\lw | m,z_s,z_l,\bp).
\end{equation}
The quantity $\tau$ is the lensing optical depth, and it contains all
of the effects of baryonic cooling on arc abundances.  (The function
$g(z_s,z_l,\bp)$ does not depend on the internal structure of the
lenses.)  In particular, the ratio of arcs in the cooling and
adiabatic simulations is simply
\begin{equation}
\frac{N_{cool}}{N_{ad}} = \frac{\tau_{cool}}{\tau_{ad}} 
	= \frac{\int dm\ (dn/dm)\ \sigma_{cool}(\lw|m)}
         {\int dm\ (dn/dm)\ \sigma_{ad  }(\lw|m)}.
\end{equation}
Our goal is to estimate this ratio.  Even this simple scenario
requires significant extrapolation from our small sample of clusters,
so we again opt for making the argument as simple as possible.  We
have then
\begin{equation}
\tau_{ad}=\int_{m_{ad}}dm\ \frac{dn}{dm}\sigma_{ad}(\lw|m).
\end{equation}
where $m_{ad}$ is the mass cutoff for the adiabatic case and we assume
a power law scaling $\sigma_{ad}=C_{ad}m^\gamma$.  For an SIS lens,
$\gamma\approx 1.3$.  For a power law halo mass function
$dn/dm=Am^{-\alpha}$ we have then
\begin{equation}
\tau_{ad}=\frac{AC_{ad}}{\alpha-\gamma-1}\frac{1}{m_{ad}^{\alpha-\gamma-1}}.
\end{equation}
A similar expression holds for $\tau_{cool}$, with $C_{cool}=\lambda
C_{ad}$ where $\lambda$ is the net boost to the lensing cross section
due to cooling, $\lambda\approx 3$.  We see then that
\begin{equation}
\frac{\tau_{cool}}{\tau_{ad}} = \lambda\left( \frac{m_{ad}}{m_{cool}}\right)^{\alpha-\gamma-1}.
\end{equation}
The second factor in parenthesis on the right hand side represents the
additional boost to the optical depth due to the lowering of the halo
mass cutoff.  To give a rough estimate for this term, note that the
cutoff mass must occur when $\lw\approx \lw_t(m)$.  We know that
$\lw_t^{cool}(m)>\lw_t^{ad}(m)$, but due to our small cluster sample,
we do not know how $\lw_t$ scales with mass in either case.  However,
we can make some progress if we assume a power law scaling between
$\lw_t$ and $m$, and further assume that cooling only changes the
amplitude of this scaling by some fixed factor $f$.\footnote{Note that
  since we do not know the scaling of $\lw_t$ with mass in either
  case, this assumption represents the simplest possible relation
  between the two scalings, and is not guaranteed to be correct.}  In
that case, the mass cutoffs $m_{ad}$ and $m_{cool}$ are related via
\begin{equation}
km_{ad}^\beta = \lw = fkm_{cool}^\beta
\end{equation}
where $\beta$ is the slope of the relation $\lw_t(m)=km^\beta$ for
adiabatic clusters.  For spherical halos, we know that $\lw_t\approx
\pi b/2R$, so in general we expect $\lw_t\propto b\propto \sigma^{1/2}
\propto m^\gamma$, and hence $\beta\approx \gamma/2$.  Finally, the
ratio $f=\lw_t^{cool}/\lw_t^{ad}\approx 1.5$ in the simulations, which
results in

\begin{equation}
\frac{\tau_{cool}}{\tau_{ad}} = \lambda f^{2(\alpha-\gamma-1)/\gamma}.
\end{equation}
For $\alpha\approx -3$ as appropriate for clusters, and $\gamma\approx
1.3$ as expected for SIS lenses\footnote{$\sigma\propto b^2\propto
  \sigma_v^4 \propto m^{4/3}$ for an SIS lens}, we find
$(\alpha-\gamma-1)/\beta\approx 1$, meaning the decrease in the halo
mass cut due to cooling can further enhance the number of arcs in the
sky by about $50\%$.  This is smaller than the net increase due to the
boost in cross section, but still far from negligible.  Moreover, for
extremely large arcs, the corresponding mass scale will be larger and
hence the mass function will be steeper, thereby enhancing the
importance of the change in the mass cutoff scale due to baryonic
cooling.


\section{Discussion and Comparison With Previous Work}
\label{sec:others}

Perhaps the most obvious point of comparison for our work is the work
of \citet[][]{puchweinetal05}, who studied the impact of baryonic
cooling on very massive ($M\gtrsim 10^{15}\ \msun$) clusters.  To the
extent that our analysis overlap, our results are in agreement and we
concur with their conclusions: baryonic cooling boosts the lensing
cross section of the most massive clusters by a factor of $\approx
2-4$, and this boost is primarily driven by the steepening of the
profile. For smaller mass clusters, this boost can be significantly
larger, which can in turn significantly increase the number of arcs
systems in the universe depending on the minimum length to width ratio
of the arcs considered.

In addition to the work by \citet[][]{puchweinetal05}, there have been
several other studies that investigated the impact of the presence of
the massive central galaxy found at the center of most clusters
\citep[][]{meneghettietal03,dalalholderhennawi04,howhite05}.  These
studies noted that when a mass model for the central cluster galaxies
is tacked on numerically simulated dark matter halos the lensing cross
section for the cluster is relatively unchanged.  This demonstrates
that the halo's response to the baryonic concentration at the center
is critically important, and strongly supports our argument that the
enhancement to the lensing cross section is driven by the associated
steepening of the halo profile.

Incidentally, we expect that the steepening of the halo profile
induced by cooling will have an additional interesting effect which we
have thus far ignored: it should increase the sensitivity of lensing
cross sections to the assumed source redshift.
\citet[][]{wambsganssbodeostriker04} used the approximation
$\mu\approx|\lr/\lt|\approx\lw$ in estimating the impact of source
redshift on lensing cross sections, and demonstrated that a broad
redshift distribution enhances the lensing optical depth.  Subsequent
work \citep[e.g.,][]{dalalholderhennawi04,lietal05} demonstrated that
the sensitivity to source redshift was overestimated in
\citet[][]{wambsganssbodeostriker04} since their approximation
$\mu\approx \lw$ is exact only for SIS profiles, with shallower
profiles resulting in less sensitivity to the source redshift
distribution. Given that baryonic cooling tends to both steepen and
circularize the halo mass profile, we expect the cooling clusters to
be more sensitive to the assumed redshift distribution.

Another interesting way in which our work can be related to previous
studies is to ask whether the general theoretical framework developed
here can shed some light on previous results.  For instance, we have
already seen that the question of the isotropy of the lensing cross
section is a strong function of the sharpness of the halo density
profile, in agreement with the arguments by
\citet[][]{dalalholderhennawi04} for flat, dark matter only profiles.
Likewise, our theoretical framework can help explain how the lensing
cross section scales with mass.  At low masses, the arc cross section
is dominated by merging arcs, and thus should quickly drop with
decreasing mass.  Conversely, above the mass cutoff, the lensing cross
section should become dominated by distortion arcs and have a much
more mild mass dependence.  This characteristic behavior of a rapidly
rising cross section at low masses transitioning into a flatter regime
seems to be at least in qualitative agreement with the cross section
mass scaling found by \citet[][]{hennawietal05}.

This same type of reasoning can be used to estimate the dependence of
lensing cross section on source radii.  For instance, consider first
the case in which the minimum length to width ratio of interest $\lw
\ll \lw_t$ where $\lw_t$ is the transition scale for a given cluster
and source size.  In this limit, the cross section is completely
distortion dominated and hence source size independent (see appendix
\ref{app:small_source} for details). If $\lw \lesssim \lw_t$,
increasing $\lw_t$ will slightly boost the image merging cross
section, so we expect the lensing cross section to increase with
increasing source size.  Conversely, at $\lw \gtrsim \lw_t$, the
length to width ratio of a source at a fixed distance $d<R$ from the
caustic will decrease with increasing source radius due since the
source then extending out to regions of lower eigenvalue ratio.  This
implies that the contour of fixed $\lw$ moves closer to the caustic,
and hence the lensing cross section will \it decrease \rm with
increasing source radius.

Based on these arguments, we expect the scaling of the lensing cross
section $\sigma(\lw)$ with source radius to be quite complicated.
This brings up the question of why should we only consider arcs as a
function of their length to width ratio?  While early works
\citep[e.g.,][]{grossmannarayan88,miraldaescude93} considered arc
abundances as a function of arc length, width, and length to width
ratio, this is no longer the case. We expect that the move towards
considering only length to width ratios came from the naive
expectation that the corresponding cross section would be source
radius independent \citep[e.g.][]{wuhammer93}.  Such a result would
imply that arc abundances could be predicted without detailed
knowledge of the radius distribution of high redshift galaxies, an
observationally challenging problem.  Unfortunately, $\sigma(\lw)$ \it
does \rm depend on source radii, so the advantage of considering
$\sigma(\lw)$ only is considerably lessened. Rather, given that the
source radius dependence of $\sigma(L)$ and $\sigma(\lw)$ are
necessarily different, it may be possible to use the abundance of
giant arcs as a function of length, width, and length to width ratio
simultaneously in order to remove source radii dependences on the
predictions.  Alternatively, it may be the case that giant arc
abundances are better suited to provide constraints on the properties
of high redshift galaxies \citep[e.g.,][]{bezecourtetal98}.  We shall
not attempt to resolve this question here.


\begin{figure}[t]
\epsscale{1.2}
\plotone{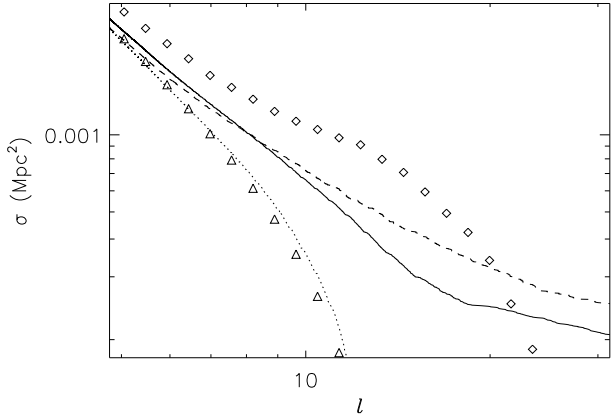}
\caption{The lensing cross section for an SIE profile with axis ratio
  $q=0.8$ is shown above with diamonds.  Also shown as triangles is
  our estimate for the distortion contribution to the lensing cross
  section obtained from our best fit model to the data (see Appendix
  \ref{app:fits}).  The solid line is obtained using the semi-analytic
  prescription of \citet[][]{fedelietal06}, while the dashed line is
  the source plane area over which the eigenvalue ratio satisfies
  $|\lr/\lt|>\lw$. Finally, the dotted line is the result from the
  semi-analytic prescription of \citet[][]{fedelietal06} after
  subtracting the amplitude at the transition scale $\lw_t$ as
  determined by our best fit.  As expected, the resulting cross
  section is in excellent agreement with our best fit estimate of the
  distortion contribution to the lensing cross section.}
\label{fig:sa_cs}
\end{figure} 


The general distinction between distortion and image merging arcs we
have introduced also has repercussions for semi-analytic calculations
of lensing cross sections.  In particular, \citet[][]{fedelietal06}
argued that one may obtain accurate cross sections using a simple
method in which the eigenvalue ratio of the inverse magnification
tensor is mapped onto the source plane, and then convolved with the
appropriate top hat filter for the source under consideration.  Since
such an algorithm explicitly ignores the possibility of image merging,
it is evident that it will underestimate lensing cross sections
whenever image merging arcs provide a non-negligible contribution to
the total lensing cross section, as shown in Figure~\ref{fig:sa_cs}.
Moreover, at fixed length to width ratio the underestimate will become
less severe as halo mass increases, an expectation that is fully
consistent with the data shown in \citet[][]{fedelietal06}.  Note that
for the massive clusters considered in \citet[][]{fedelietal06}, we do
indeed expect excellent agreement between the semi-analytic estimates
and the true lensing cross section.

As a final application of our results, we consider now the \it
multiplicity ratio, \rm the number of clusters with multiple arcs over
the number of clusters hosting a single arc.  This ratio was first
introduced by \citet[][]{gladdersetal03} to argue that there must be a
population of super lenses: clusters of ordinary mass that for some
unknown reason, most likely projection effects, have extraordinarily
large cross sections relative to other clusters of comparable mass.
This raises the interesting question of how does baryonic cooling
affect the multiplicity ratio for a fixed cosmology.

We estimate the impact of baryonic cooling on the multiplicity ratio
as follows: the probability $P(2^+|1^+)$ that a cluster with at least
one arc host two or more arcs is given by
\begin{equation}
P(2^+|1^+)=\int d\mu P(2^+|\mu)P(\mu|1^+)
\end{equation}
where $P(2^+|\mu)$ is the probability of observing more than one arc
in a halo where the expected number of arcs is $\mu$.  $P(\mu|1^+)$ is
the probability that a halo hosting at least one arc has an
expectation number of arcs $\mu$.  By Bayes's theorem,
\begin{equation}
P(\mu|1^+)=\frac{P(1^+|\mu)P(\mu)}{P(1^+)}
\end{equation}
and the probability that a halo host at least one arc is 
\begin{equation}
P(1^+) = \int d\mu\ P(1^+|\mu) P(\mu).
\end{equation}
Putting it all together we find
\begin{equation}
P(2^+|1^+) = \frac{\int d\mu\ (P(2^+|\mu)P(1^+|\mu)P(\mu)}{\int d\mu\ P(1^+|\mu)P(\mu)}.
\label{eq:mult}
\end{equation}

To compute the multiplicity ratio, we assume Poisson statistics for
$P(N|\mu)$, the probability of a halo hosting $N$ arcs given its
expectation value $\mu$.  For $P(\mu)$ we use the fact that for a halo
of mass $m$, $\mu \propto \sigma \propto m^\gamma$, so that $P(\mu)
\propto \mu^{-\alpha+\gamma-1}$ where $\alpha$ is the slope of the
halo mass function at the lensing scale.  Finally, the integrals of
the above expression are truncated at a lower limit $\mu_0$ where we
naively expect $0.1\lesssim \mu_0 \lesssim 1$.  Assuming $\mu_0$ is
the expected number of giant arcs due to clusters at the cutoff mass
$m_{ad}$ or $m_{cool}$, we obtain
\begin{equation}
\frac{(\mu_0)_{cool}}{(\mu_0)_{min}} =
\frac{\sigma_{cool}(m_{cool})}{\sigma_{ad}(m_{ad})} \approx \lambda/f^2\approx 1.5
\end{equation}
where to obtain a numeric value we used $\gamma\approx 1.3$ as
appropriate for an SIS.  Thus, even though cooling lowers the lensing
mass cutoff, it may slightly increase the effective cutoff in $\mu$,
and thus slightly enhance the multiplicity ratio.  Perhaps more
important, however, is to recognize the fact that our above argument
shows that the natural expectation value for the multiplicity ratio is
in the range $\approx 0.2-0.6$ as can be directly computed from
equation \ref{eq:mult}, and that this ratio should be fairly robust to
details of the halo profile and cosmology, as observed by
\citet[][]{howhite05}.


\section{Summary and Conclusions}
\label{sec:summary}

Using simple analytic halo density profiles, we have investigated how
lensing cross sections depend on various halo parameters.  In
particular, we found that the lensing cross section has three distinct
contributions corresponding to arcs formed through image distortion
only, merging of two images, and merging of three images.  Moreover,
$\sigma(\lw)$ exhibits a knee at the scale which the cross section
goes from being distortion dominated to image merging dominated, and
the latter component falls exponentially fast with increasing length
to width ratio.

We then proceeded to investigate how these various contributions to
the lensing cross section depend on halo properties.  In particular,
we found that while the image merging contribution to the cross
section increases with increasing halo ellipticity, the distortion
contribution remains relatively constant, except for the fact that it
gets truncated at a smaller transition scale $\lw_t$.  In other words,
$\lw_t$ decreases with increasing ellipticity.  We also investigated
the impact of the slope of the profile on the lensing cross section,
and found that at fixed Einstein radius, steeper profiles result in
more effective lenses, and more circular distribution of arcs about
the cluster center.  We also noted that the ratio of cusp arcs to fold
arcs appears to be very sensitive to the slope of the density profile.

Based on these observations, we argued that the probability for
finding arcs of a given length to width ratio should quickly rise with
mass below some mass threshold $M_{min}$, and that above this mass cut
the scaling should become much flatter.  We also discussed how the
distinction between arcs formed through image merging and arcs formed
through distortion leads to a non-trivial dependence of the lensing
cross section on the assumed source radius.

We also investigated the impact of baryonic cooling on the lensing
cross section, and found that the steepening of the halo density
profile in response to the central baryon condensation enhances the
lensing cross section at low $\lw$ by about a factor of three, and
lowers the mass threshold $M_{min}$ for halos to become effective
lenses.  The latter effect helps further increase arc abundances by
about $50\%$.  Moreover, cooling clusters had a significantly larger
probability of forming fold arcs than adiabatic clusters, and their
lensing cross section was much less dependent upon the line of sight
projection axis than adiabatic clusters.  Both of these effects are
also explained by the steepening of the halo profile brought about by
the contraction of the halo in response to the condensation of cooling
baryons at the center.  Finally, we also argued that substructures did
not appear to significantly enhance lensing cross sections, and that
the multiplicity ratio, i.e.  the number of lenses that exhibit
multiple arcs, is only moderately affected by the details of the mass
distribution of the lenses.

Given the sensitivity of arc abundances to baryonic processes, it is
difficult to see this observable becoming a tool for precision
cosmology in the near future.  Conversely, as the range of
cosmological parameters get narrower from other observations, this
same sensitivity can be used to probe cluster properties and/or the
properties of high redshift galaxies.  Concerning this last
possibility, it seems clear that a systematic analysis similar to the
one carried here aimed at understanding how various source properties
affect lensing cross section would be a fruitful endeavor which we
intend to carry out.

\it Acknowledgments: \rm E.R. would like to thank Scott Dodelson for a
careful reading and helpful suggestions throughout the development of
this work.  This work was supported in part by the Kavli Institute for
Cosmological Physics through the grant NSF PHY-0114422.  DN is
supported by the Sherman Fairchild Postdoctoral Fellowship at Caltech.
AVK is supported by the National Science Foundation (NSF) under grants
No. AST-0206216 and AST-0239759 and by NASA through grant NAG5-13274.
The authors would also like to thank KICP Thunch, where many
conversations related to this topic took place.


\bibliography{ms}  


\appendix
\section{Lensing Algorithm}
\label{app:algorithm}

In this section, we describe the details of our lensing algorithm,
including various procedures to secure a 10\% accuracy in the cross
section and several techniques to improve the speed of the algorithm.


\subsection{Source Placement, and Image Identification and Classification}

Having defined the lens and source plane grids, we define the placement
grid resolution $r_p$ via $r_p=R/N$ with $N=5$ as our default value,
which is fine enough to adequately resolve the source plane region where
sources touch the caustic of the lensing cluster.  Now, for our most
massive systems, the placement grid may easily have $\gtrsim
2000$ pixels per side with $\lesssim 1\%$ of these pixels producing
giant arcs.  We therefore seek an efficient algorithm for source
placement and image processing.

We proceed as follows.  We begin by first placing sources on all
grid pixels that map onto a region where $\lr/\lt > 0.9\lwmin$ and
iteratively place sources along the edges of the above region until
the placement of additional source do not result in new arcs. We
record the position, length, and width of every image generated,
as well as the corresponding source position.

The images of an individual sources are processed as follows.  First,
a list of all lit (not necessarily contiguous) image pixels is made.
A random pixel is selected, and we move along the $x$ and $y$ axis
until we hit the edge of the image, collecting all lit pixels as we
go along.  For each pixel collected, we then move in the orthogonal
direction, and iterate until no new pixels are collected.  We also
keep track of edge pixels.  The output of this procedure is thus a
list of each of the disjoint images along with the edges of each of
these images.  We note that the procedure described above is much more
efficient than the simple neighboring pixel search which is typically
employed.

Consider now a single arc.  To compute the length and width of the
arc we first identify the image pixel containing the center of the
source (call it $C$), the pixel farthest from it (call it $F_1$),
and the pixel $F_2$ farthest from $F_1$.  The length of the image
is defined to be that of the circular arc defined by the points
$F_1\,C\,F_2$ plus one pixel unit, the latter being a geometrical
correction introduced by \citet[][]{puchweinetal05}. Finally, we
define the width of the image via $\pi LW=4A$ where $A$ is the area
of the image, $L$ is its length, and $W$ its width. Our definition
is motivated by the fact that for an ellipse, $\pi LW=4A$.


\begin{figure}[t]
\epsscale{0.8} \plotone{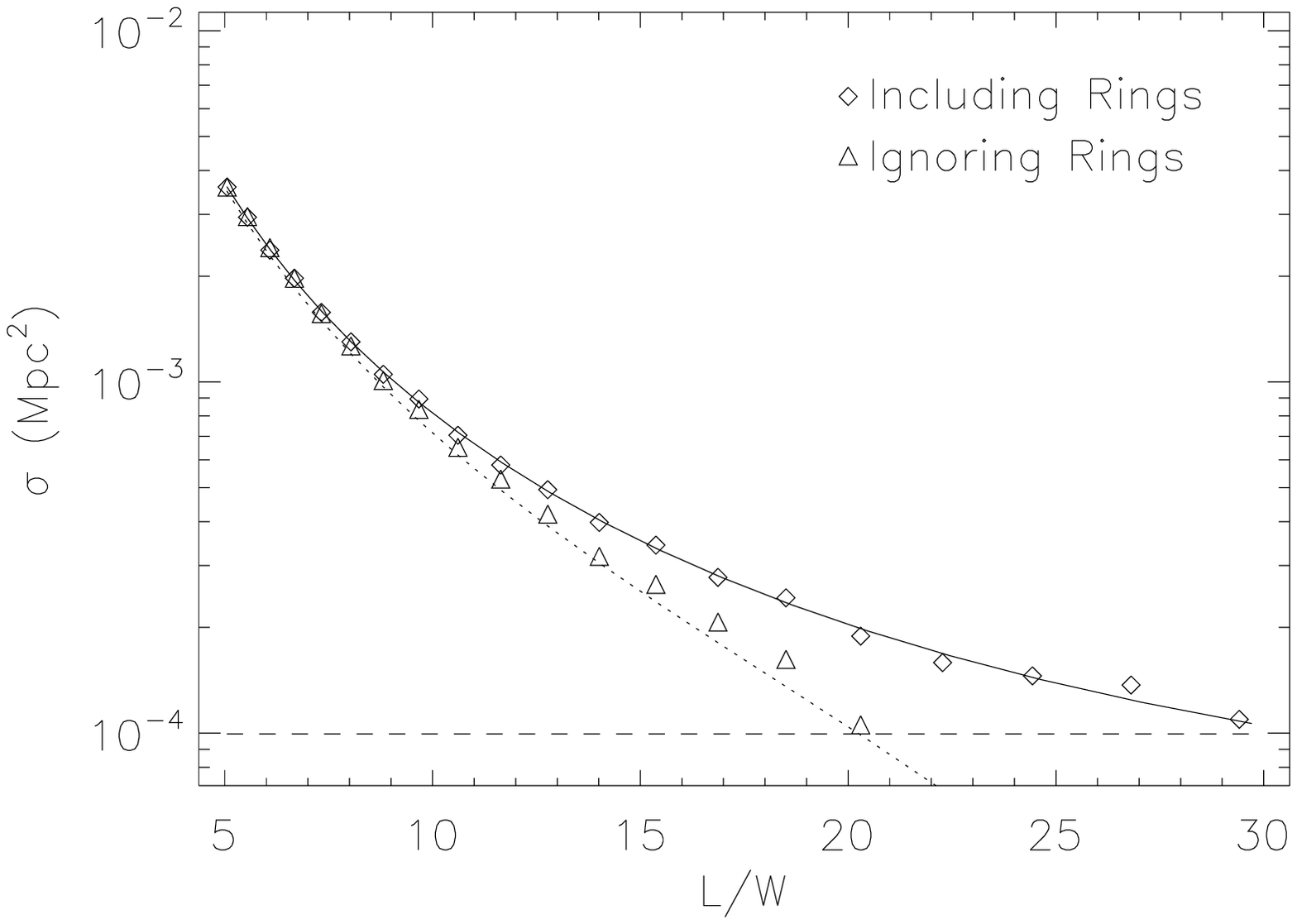}
\caption{Effects of ring images on the lensing cross section $\sigma(\lw)$.
The {\it triangles} and {\it diamonds} points show $\sigma(\lw)$ as a
function of L/W with and without ring images for an SIS profile with
velocity dispersion $\sigma_v = 10^3 \kms$ and lens and source
redshifts $z_l=0.3$ and $z_s=1.5$.  This shows that failure to treat
rings as such can result in biased lensing cross sections of
circularly symmetric profiles.  The {\it solid} line fit is our
analytic fit composed of a power-law distortion component ({\it dotted
line}) plus a constant contribution corresponding to the cross section
for forming rings ({\it dashed line}).  Note that the total cross
section with rings consists of the cross section without rings plus a
constant contribution corresponding to the cross section for forming
rings, $\pi R^2$.}
\label{fig:ring_selection}
\end{figure} 


\subsection{Circular Profiles and Formation of Rings}

There is one special case that our algorithm does not treat properly:
that of sources that produce full Einstein rings, since for the latter
$A\approx 2\pi LW$ where $W$ is the width of the ring. The obvious way
to correct for this is to classify images as arcs or rings, but this
can be laborious.  Instead, we have opted for a statistical approach.
The basic idea is simple: assuming rings are formed, the width
distribution of the longest arcs in the sample will be bimodal with a
large gap of about a factor of two corresponding to the transition
between disjoint arcs and rings.  We can thus identify rings by
searching for this gap, and then identifying all of the wide ``arcs''
as rings. Operationally, we identify all arcs with lengths and widths
larger than some length and width cuts $L_{cut}$ and $W_{cut}$ as
rings, where we set $L_{cut}$ to be 80$\%$ of the length of the widest
arc and a width cut $W_{cut}$ to be the median width of all arcs with
$L<L_{cut}$.  Figure~\ref{fig:ring_selection} shows the effects of
rings on the lensing cross section $\sigma(\lw)$.  It shows that
failing to treat rings properly  resulting in an underestimate of the
giant arc cross section for high $\lw$ values since rings are being
misidentified as relatively small (low $\lw$) arcs.  To correct
for this, in this work we
simply assign to all rings a length to width ratio $\lw=$max$(\lw)$
where max$(\lw)$ is the maximum length to width ratio of all other arcs.


\subsection{Selection Function: Choosing Tangential Arcs}

Arcs can form in one of two ways: through strong distortion along the
tangential direction, or through the merging of different images of a
single source.  This is true for both tangential and radial
arcs. Therefore, the cross section $\sigma(\lw)$ generally consists of
four distinct contributions.  However, we have chosen to focus on
tangential arcs only, since tangential arcs are much more prominent
and easily identifiable than radial arcs.  Observationally, this has
the advantage that the purity and completeness of a tangential arc
sample will be higher than that of a radial arc sample.  It is
therefore important to identify and remove radial arcs when computing
the cross section for tangential arcs.

We have chosen to identify radial arcs through a statistical procedure
similar to the one used to identify rings.  In particular, we exploit
the fact that radial arcs tend to be not just magnified along the
radial direction, but also {\it demagnified} along the tangential
direction.  In other words, the radial arcs create a distinct
population of images branching out from the main population and
reaching the low L and low W part of the plane.  We therefore tag all
arcs with $L<L_{cut}$ and width $W<W_{cut}$ as radial arcs.  The cuts
are defined through the following algorithm: first we find the median
width of all arcs above the minimum length to width cut $\lwmin$,
which is our first estimate for $W_{cut}$.  We then lower $W_{cut}$ in
steps of $0.5r_p$ where $r_p$ is the placement grid resolution until
the increase in the number of arcs with $W>W_{cut}$ is less than
$5\%$.  The length cut $L_{cut}$ is defined as
$L_{cut}=W_{cut}\lwmin$ where $\lwmin$ is the minimum length
to width ratio of arcs considered.  We found these cuts cleanly
separate the tangential and radial branches.


\section{Validity of the Small Source Approximation}
\label{app:small_source}

One of the key steps in our lensing finding algorithm is selecting all
source placement grid pixels where the eigenvalue ratio $\lr/\lt$ is
larger than 90$\%$ of the minimum length to width ratio considered.
This selection is based on the expectation that for infinitely small
sources, the eigenvalue ratio $\lr/\lt$ will be precisely the length
to width ratio $\lw$ of the resulting image.  A natural question that
arises then is to what extent is this small source approximation
valid.

\begin{figure}[t]
\epsscale{0.8}
\plotone{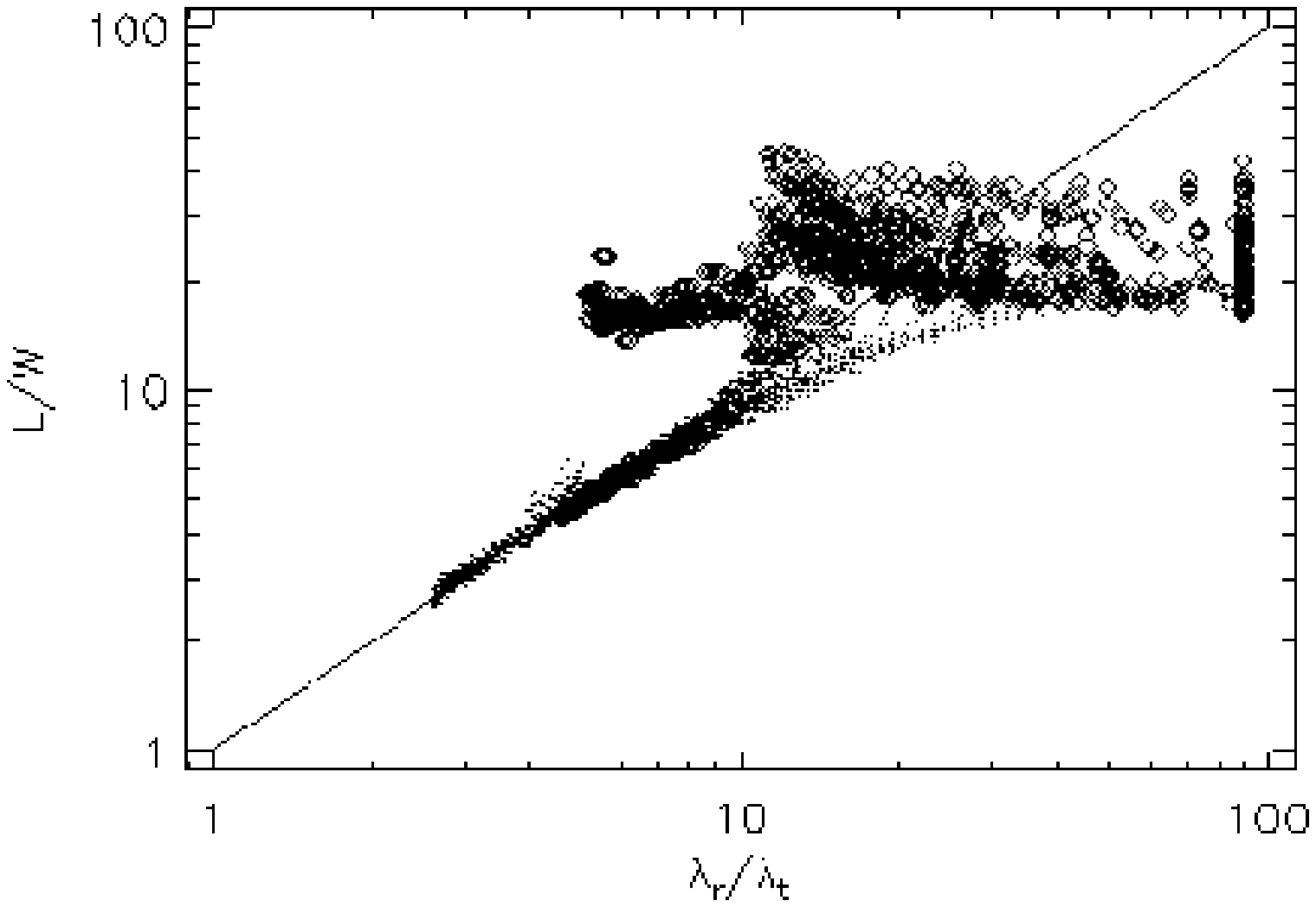}
\caption{$\lw$ vs. eigenvalue ratio. 
The plot is constructed as follows: we begin with an
SIE profile with a velocity dispersion $\sigma_v=10^3 \kms$ and an
axis ratio $q=0.8$.  For each source we place on the source plane, we
find its largest arc, and the eigenvalue ratio $\lr/\lt$ at the center
of said arc.  To keep the $x$-axis small, we set $\lr/\lt=90$ for sources
in which $\lr/\lt>90$.
Dots represent sources which do not touch the tangential
caustic, whereas diamonds correspond to sources that do touch the
tangential caustic.  It is evident from the figure that the linear
approximation $\lw = \lr/\lt$, shown here with the solid line, is a
good one even for strongly distorted images, but that the agreement
fails if the source touches a caustic.  This reflects the fact that
the length of these arcs is set by the merging of two images.}
\label{fig:r_vs_egvr}
\end{figure}

Figure~\ref{fig:r_vs_egvr} illustrates the relation between $L/W$ and
$\lr/\lt$ for an SIE model with $\sigma_v = 10^3 \kms$ and $q=0.8$.
Each point in the figure represents a source that does not touch the
lens's caustic, while sources that touch the tangential caustic are
shown as diamonds.  This shows that the agreement between $\lr/\lt$
and $\lw$ is excellent provided the source in question does not touch
the lens's caustic.  This strongly suggests that the generalization of
the eigenvalue ratio $\lr/\lt$ estimate for $\lw$ to elliptical
sources proposed by \citet[][]{keeton01b} should be accurate for
distortion arcs, as was indeed found by \citet[][]{fedelietal06}.


\section{Fitting Function for Lensing Cross Sections}
\label{app:fits}

In this section, we develop an analytic fitting function to the
lensing cross sections useful for both analytic and numerical models
of galaxy clusters.  The idea is based on the realization that at low
$\lw$, the lensing cross section curve is dominated by distortion arcs
is well fit by a power law, whereas at large $\lw$ ratios, the cross
section is dominated by image merging arcs, and is better fit by an
exponential fall off.  So, the total cross section $\sigma(\lw)$ can be
expressed as a sum of these two components:
\begin{equation}
\sigma(\lw)=\sigma_{distortion}(\lw)+\sigma_{merging}(\lw)
\end{equation}
where
\begin{equation}
\sigma_{distortion}(\lw) = \left\{\begin{array}{ll}
  \sigma_0\left[\left(\frac{\lw_0}{\lw}\right)^\nu-
  \left(\frac{\lw_0}{\lw_{d,max}}\right)^\nu\right] & \mbox{if}\ \lw<\lw_{d,max}, \\
  0 & \mbox{if}\ \lw>\lw_{d,max}, 
  \end{array} \right.
\end{equation}
and
\begin{equation}
\sigma_{merging}(\lw) = \left\{\begin{array}{ll} 
  \sigma_c &  \mbox{if}\  \lw<\lw_t, \\
  \sigma_c \exp\left[ -(\lw-\lw_t)/\lw_s \right]  & \mbox{if}\ \lw>\lw_t, 
  \end{array} \right.
\end{equation}
In the above expressions, $\lw_{d,max}$ is the maximum length to width
ratio of distortion arcs, and $\lw_t$ is the minimum length to width
ratio of image merging arcs.  The subscript $c$ denotes `caustic', in
that image merging sources are those that touch the lens's caustic.

While $\lw_{d,max}$ tends to be difficult to resolve\footnote{We often
found that $\lw_{d,max}\lesssim \lw_t$, and $\lw_{d,max}$ tends to be
difficult to resolve since the cross section at these scales is
dominated by the image merging contribution.}, we may simply set
$\lw_{d,max}=\lw_t$ in the above equation.  This results in the fitting
function:
\begin{equation}
\sigma(\lw)=\left\{\begin{array}{ll}
	\sigma_0\left[\left(\frac{\lw_0}{\lw}\right)^\nu-
		\left(\frac{\lw_0}{\lw_t}\right)^\nu\right]
		+\sigma_c & \mbox{if}\ \lw\leq \lw_t \\
	\sigma_c \exp\left[ -(\lw-\lw_t)/\lw_s \right] & \mbox{if}\ \lw>\lw_t
	\end{array} \right.
\label{eq:ansatz}
\end{equation}
Our ansatz in equation \ref{eq:ansatz} provides an \it excellent \rm
fit to the data whenever the distortion and image merging
contributions can be cleanly separated.\footnote{This is essentially
all cases we considered with the exception of most $\beta=0.6$ models,
and numerical clusters where the presence of substructures severely
distort the shape of the lensing cross section.}  All fits were done
with logarithmic sampling on $\lw$, and assuming an arbitrary $10\%$
error bar at each point.\footnote{These do not represent real errors,
which will in reality be strongly correlated, and are simply used to
define a best fit model.  Consequently, the $\chi^2$ for these fits
cannot be used to judge whether the fits are statistically acceptable
or not. We choose our particular fitting scheme since in the absence
of a clear way to assign error bars to our measurements, simplicity
was the next obvious criteria for choosing a fitting algorithm.}
Furthermore, the $\chi^2$ surface typically exhibits only one deep
minima, implying that degeneracies between the various parameters do
not exist when the two components of the lensing cross section are
clearly distinct.  


\begin{figure}[t]
\epsscale{0.8}
\plotone{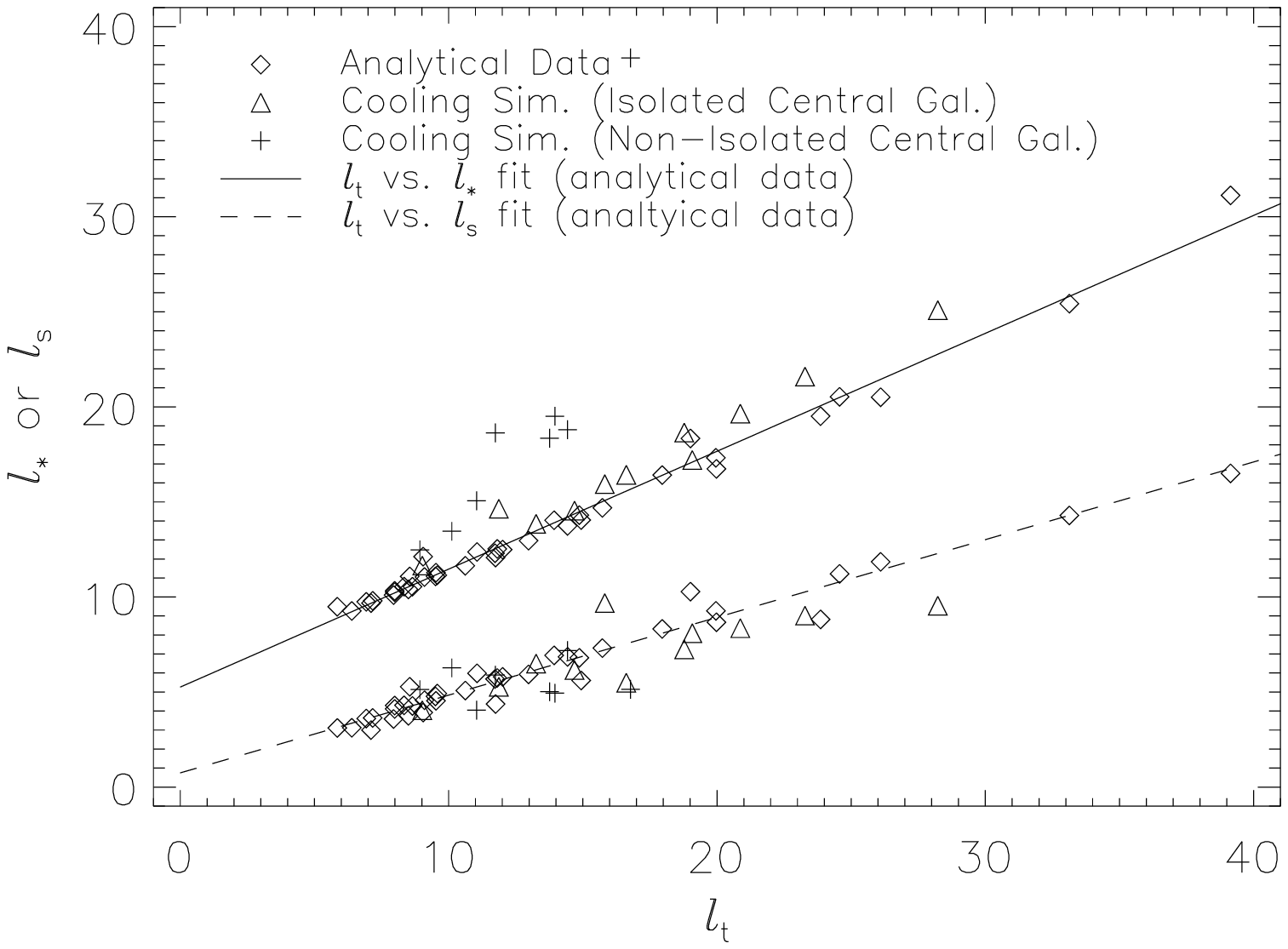}
\caption{Correlations between the fitting
parameters of equation \ref{eq:ansatz}.  In addition to the
correlation between $\lw_t$ and $\lw_s$ shown by a dashed line, there
is an additional correlation between $\lw_t$ and the parameter $\lw_*$
defined through $\sigma_0 (\lw_0/\lw_*)^\nu = \sigma_c$.  This latter
correlation is shown with the solid line.  We have used a least square
fitting to define the correlation parameters, and have considered only
data in which the cross section function exhibited a clear transition
scale, with the added provision that the image merging contribution to
the cross section by dominated by fold arcs.  Note that the $\lw_*$
values have been displaced upwards by five units for illustration
purposes.  Also shown above are the best fit values obtained from
$x$, $y$, and $z$ projections of simulated clusters including cooling
and star formations, divided into clusters with a relatively isolated
central component (triangles) and clusters with large amounts of
substructure near the central density peak (crosses).  Note that the
best fit relation obtained from our analytic profiles appears to be
satisfied by numerical clusters as well.}
\label{fig:par_correlations}
\end{figure} 


Note that every parameter has a concrete physical meaning, but these
parameters are clearly correlated.  For example, we expect the
parameters $\lw_t$ and $\lw_s$ to be correlated, since $\lw_s$ is a measure
of how quickly the length to width ratio of an arc decreases as the
source moves away from the lens's caustic, and $\lw_t$ is the minimum
length to width ratio such a source can have while still touching the
caustic.  Another interesting correlation is between the parameter
$\lw_*$ and $\lw_t$ where $\lw_*$  is defined via
\begin{equation}
\sigma_c = \sigma_0(\lw_0/\lw_*)^\nu.
\label{eq:rstar}
\end{equation}
Thus, $\lw_*$ is the length to width ratio at which the image merging
cross contribution to the cross section is as large as the image
distortion contribution.  Figure~\ref{fig:par_correlations} shows that
a clear and tight correlation exists between $\lw_t$ and $\lw_s$ (dashed
line) and between $\lw_t$ and $\lw_*$ (solid line) for both analytical and
simulated cluster data spanning a large range of ellipticities, halo
profiles, and Einstein radii.  The best fit lines to the $\lw_t$--$\lw_s$
and $\lw_t$--$\lw_*$ correlations are given by
\begin{eqnarray}
\lw_s & = & 0.748+0.409\lw_t, \label{eq:rs_rt_corr} \\
\lw_* & = & 0.260+0.620\lw_t. \label{eq:rstar_rt_corr} 
\end{eqnarray}
Note that in Figure~\ref{fig:par_correlations}, the $\lw_*$ values have
been displaced upwards by five units for illustration purposes.

The presence of strong correlations among fit parameters suggests
that five parameters are not needed to adequately fit the data;
three parameters will suffice, namely $\sigma_0$, $\nu,$ and $\lw_t$.
It turns out that there is a tight correlation between $\nu$ and
$\lw_t$ as well, further suggesting that lensing cross section
curves $\sigma(\lw)$ may, in general, be characterized through
only {\em two} parameters, an amplitude $\sigma_0$ and a transition
scale $\lw_t$.  This is intriguing since the minimal number of
parameters to describe a halo are three: an Einstein radius, an
ellipticity, and slope profile $\beta$, which raises the interesting
question of whether it is possible for two different halos to have
the exact same lensing cross section function $\sigma(\lw)$.

\end{document}